# Effect of Genetic Variation in a Drosophila Model of Diabetes-Associated Misfolded Human Proinsulin

**Bin Z. He[*,1,2], Michael Z. Ludwig[*], Desiree A. Dickerson[*], Levi Barse[*], Bharath Arun[*], Soo-Young Park[†], Natalia A. Tamarina[†], Scott B. Selleck[§], Patricia J. Wittkopp[§§], Graeme I. Bell[†,‡], Martin Kreitman[*,2]**

[*] Department of Ecology and Evolution, The University of Chicago, Chicago, IL 60637

[†] Department of Medicine, The University of Chicago, Chicago, IL 60637

[‡] Department of Human Genetics, The University of Chicago, Chicago, IL 60637

[§] Department of Biochemistry and Molecular Biology, The Pennsylvania State University, University Park, PA, 16802

[§§] Department of Ecology and Evolutionary Biology, and Department of Molecular, Cellular, and Developmental Biology, University of Michigan, Ann Arbor, MI 48109

[1] Current Address: FAS Center for Systems Biology, Harvard University, 52 Oxford Street, Cambridge, MA 02138

Supporting data is available online at http://openwetware.org/wiki/Kreitman:Publications

[2] Corresponding authors:
Martin Kreitman
Mailing address: Department of Ecology and Evolution, The University of Chicago, 1101 E 57th Street, Chicago, IL 60637-1573.
Phone: +1 773 702 1222. Fax: +1 773 702 9740.
Email: martinkreitman@gmail.com.

Bin Z. He
Mailing address: FAS Center for Systems Biology, Harvard University, 52 Oxford Street, Cambridge, MA 02138
Phone: +1 312 550 8421. Fax: +1 617 496 5425
Email: binhe@fas.harvard.edu

## ABSTRACT

The identification and validation of gene-gene interactions is a major challenge in human studies. Here, we explore an approach for studying epistasis in humans using a *Drosophila melanogaster* model of neonatal diabetes mellitus. Expression of mutant preproinsulin, $hINS^{C96Y}$, in the eye imaginal disc mimics the human disease activating conserved cell stress response pathways leading to cell death and reduction in eye area. Dominant-acting variants in wild-derived inbred lines from the Drosophila Genetics Reference Panel produce a continuous, highly heritable, distribution of eye degeneration phenotypes. A genome-wide association study (GWAS) in 154 sequenced lines identified 29 candidate SNPs in 16 loci with $P < 10^{-5}$ including one SNP in an intron of the gene *sulfateless* (*sfl*) which exceeded a conservative genome-wide significance threshold of $P = 0.05$ level ($-\log_{10} P > 7.62$). RNAi knock-downs of *sfl* enhanced the eye degeneration phenotype in a mutant-hINS-dependent manner. *sfl* encodes a protein required for sulfation of the glycosaminoglycan, heparan sulfate. Two additional genes in the heparan sulfate (HS) biosynthetic pathway (tout velu, *ttv* and brother of tout velu, *botv*) also modified the eye phenotype, suggesting a link between HS-modified proteins and cellular responses to misfolded proteins. Finally, intronic variants marking the QTL were associated with decreased *sfl* expression, a result consistent with that predicted by RNAi studies. The ability to create a model of human genetic disease in the fly, map a QTL by GWAS to a specific gene (and noncoding variant), validate its contribution to disease with available genetic resources, and experimentally link the variant to a molecular mechanism, demonstrate the many advantages Drosophila holds in determining the genetic underpinnings of human disease.

## INTRODUCTION

Limitations imposed by human subject research can be overcome by investigating models of human disease in experimental organisms. *Drosophila* routinely provide genetic insights relevant to human biology and disease, owing to the deep conservation of fundamental cellular and developmental processes. Motivated by the success of this approach, we constructed a fly model of protein misfolding disease, by creating a transgene of a diabetes-causing, human mutant preproinsulin (hINS$^{C96Y}$) that could be expressed in the eye imaginal discs and other developing tissues (Park et al., 2013). The misfolded proinsulin protein causes the loss of insulin-secreting pancreatic beta cells and diabetes in humans and mice (Støy et al., 2007). When misexpressed in the *Drosophila* eye imaginal disc, it disrupts eye development, resulting in a reduced eye area in adult flies (Park et al., 2013).

To investigate the effect of genetic variation in this model of a human disease, we crossed the transgenic line bearing the mutant preproinsulin and an eye-specific Gal4 driver (GMR>>hINS$^{C96Y}$) with 178 lines from the Drosophila Genetics Reference Panel (DGRP) (Mackay et al., 2012) . The resulting F1 offspring exhibited extensive and highly heritable variation in the extent of eye degeneration (Park et al., 2013). The nearly continuous distribution of eye degeneration phenotypes among the lines suggested a polygenic basis for this genetic background variation.

Drosophila's many favorable attributes for mapping quantitative trait loci (QTL) — a high density of common variants, relatively little population subdivision, a decay of linkage disequilibrium (LD) over a scale of only 100's of bp, controlled crosses allowing repeat measurements, and excellent resources for confirmatory genetics — allowed us to identify a variant in the heparan sulfate biosynthesis pathway gene, *sulfateless* (*sfl*), and then validate the gene by genetic analysis. Moreover, studies of two other genes in the HS biosynthetic pathway, *tout-velo* (*ttv*) and *brother of tout-velo* (*botv*), showed a similar effect implicating the HS-modified proteins, or proteoglycans (HSPG), in the response to misfolded proteins.

## MATERIALS AND METHODS

### *Drosophila stocks and crosses*

The {GMR-Gal4, UAS-hINS$^{C96Y}$} line was generated by crossing the GMR-Gal4 line (Stock #1104, Bloomington Stock Center) with the UAS-hINS$^{C96Y}$ line (Park et al., 2013), and obtaining the recombinant 2nd chromosome, which was balanced over CyO. DGRP lines were obtained from the Bloomington stock center. RNAi lines against *sfl* (GD5070), *ttv* (GD4871), *botv* (GD37186) were from the Vienna *Drosophila* RNAi center. Mutant lines for *ttv* (*ttv*$^{681}$) and *botv* (*botv*$^{510}$) were described previously (Ren et al., 2009).

### *Eye area measurement*

All crosses were reared at 25°C. Total eye area was measured as described in (Park et al., 2013). At least 10 images (independent flies) passing the quality check were collected for each cross. Raw data is available on the authors' website (http://openwetware.org/wiki/Kreitman:Publications).

### *Principal Component Analysis*

The whole-genome SNP dataset for the 154 DGRP lines used for GWAS (see **Table S1** for the list of line numbers) was downloaded from the DGRP website (http://dgrp.gnets.ncsu.edu/). To detect population structures, 900K SNPs (after LD pruning using PLINK v1.07, with parameter --indep-pairwise 50 5 0.5) were used to identify the top 15 principal components (PCs) (SmartPCA software in EIGENSOFT v3.0, no outlier exclusion). We tested for the presence of population structure in the sample, a possible confounding source of association in GWAS, by measuring the correlation between the hINS$^{C96Y}$ phenotype (line mean) and projection length in the direction of the top five principle components in each DGRP line.

### *Genome wide association*

The mean eye area of 154 DGRP lines crossed to the hINS$^{C96Y}$ line was regressed on each SNP with a minor allele frequency MAF > 5% (PLINK 1.07, quantitative trait mode). 2,106,077 autosomal and 324,253 SNPs on the X chromosome were tested. The F1 males in this cross received their X chromosome from the transgene-containing strain. The identity by descent of this X chromosome allowed us to test whether the X-linked

SNPs in the DGRP sample conformed to a null distribution assuming no association. This was tested and confirmed in quantile-quantile (Q-Q) plot analysis. Because an estimate of the total number of independent SNPs genome-wide does not exist for the fly, we adopted two thresholds to identify candidate SNPs on the autosomes. The first is a Bonferroni corrected threshold at $P = 0.05$ level ($-\log_{10} P > 7.62$). This is conservative because it assumes all tests are independent while the number of independent SNPs must be much smaller than the total number tested. The second is an arbitrary threshold of $P < 10^{-5}$ (or $10^{-6}$) as suggested by Mackay et al. (Mackay et al., 2012). We estimated the false discovery rate (FDR) associated with a $P < 10^{-5}$ threshold by using X chromosome SNPs as negative controls and assuming an equal FDR between X chromosome and autosomes. We also compared the number of QTLs passing this threshold with an empirical null distribution by randomizing the phenotype relative to the genotype 2,000 times and carrying out GWAS on each of the permuted datasets.

***Conditional analysis using sfl intronic SNPs as covariates***

To identify possible secondary associations in *sfl* or elsewhere in the genome independent of the intronic QTL variants in *sfl*, we fit a linear model with the most significant one, a 18 bp /4 bp insertion/deletion polymorphism, as a covariate. This analysis was performed either within the *sfl* locus or genome-wide, and in each case the p-values were corrected for multiple testing using Bonferroni's method.

***Expression of sfl and CG32396***

Expression profiles in adult tissues were assessed using data from FlyAtlas (Chintapalli et al., 2007) and modENCODE (Roy et al., 2010). To assay expression in the eye imaginal discs, we isolated total RNA from 10 pairs of discs from 3rd instar larvae. Individual larva were sexed and dissected in 1X phosphate buffer saline (PBS); the eye portions of the eye-antennal disc were collected and the isolated discs immediately dissolved in 300 μl Trizol (Invitrogen). Total RNA was extracted according to the manufacturer's instructions. cDNA libraries were constructed using (dT)20 primers after DNase I treatment (Invitrogen). Real time quantitative PCR was performed with primer pairs targeting either *sfl* or CG32396, with expression of the gene *rp49* as an endogenous reference (SYBR-Green assay). Primers used for qRT-PCR are listed in **Table S2**.

### *RNAi and validation studies*

All RNAi lines were originally from the Vienna Drosophila RNAi Center as P-element insertion lines on a co-isogenic w1118 background. Each RNAi line was first tested to determine whether it alone had an effect on eye development by crossing it to GMR-Gal4 and comparing the eye area of the F1 males (or females) to the control cross between w1118 and GMR-Gal4. In all crosses, GMR-Gal4 was used as the maternal parent. To test its effect on the $hINS^{C96Y}$-induced eye degeneration phenotype, the RNAi line was crossed to the GMR>>$hINS^{C96Y}$ line (used as maternal parent), so that both $hINS^{C96Y}$ and the RNAi constructs are driven by GMR-Gal4. The resulting phenotype was compared to the cross between $hINS^{C96Y}$ females and w1118 males. At least 10 individual flies were measured per cross and a t-test was used to determine significance at 0.05 level with multiple testing correction. For mutant lines, GMR-Gal4 was replaced with w1118 in the first test and used as a control. The same scheme was used for the second test. It is worth noting that because the mutants were tested in heterozygous states, only dominant interaction with $hINS^{C96Y}$ will be revealed.

### *sfl expression studies*

Six lines carrying the 18 bp indel allele and eight carrying the 4 bp allele were randomly chosen and paired to form 15 crosses (**Figure S1**A). Three sets of ten $3^{rd}$ instar wandering larvae were collected from each cross and dissected in 1X PBS to isolate eye imaginal discs. RNA isolation and cDNA library preparation are the same as described above. Genomic DNA was extracted from adult flies from the same cross. Because the 18 bp/4 bp polymorphism is in the intron of *sfl*, a SNP in the cDNA was identified that could be used to distinguish the two alleles in each cross (**Figure S1**B). Four such SNPs were chosen and pyro-sequencing assays were designed (primers listed in **Table S2**). Pyro-sequencing was performed as previously described (Wittkopp, 2011). Briefly, each of the three cDNA and one gDNA sample per cross was analyzed by pyrosequencing in four replicate PCR amplifications to determine relative expression.

The ratio in genomic DNA analysis was used to account for amplification bias. The resulting 12 ratios were first log2 transformed and analyzed using ANOVA according to the model $y_{ij} = \boldsymbol{\alpha} + \boldsymbol{L_i} + \boldsymbol{\varepsilon_{ij}}$, where $\alpha$ is the estimate of the relative expression ratio, which is expected to be significantly different from zero when the two alleles are differentially expressed; $L_i$ is a random effect term for the biological replicates ($i$ = 1,2,3). For 13 of the 15 crosses the p-value > 0.1; for these crosses the data were fit a with a reduced ANOVA model $y_i = \boldsymbol{\alpha} + \boldsymbol{\varepsilon_i}$, from which the estimate and the 95% confidence interval for the ratio of expression ($\alpha$) were calculated. In the two cases where the random effect term was nominally significant ($P < 0.1$), a linear mixed-effect model was fit using the lme package in R to obtain an estimate and 95% confidence interval for the same ratio.

## RESULTS

### *Effect of natural variation on hINS$^{C96Y}$-induced eye phenotype*

We crossed the transgenic fly line (w; P{GMR-Gal4}, P{UAS-hINS$^{C96Y}$ }/CyO) as the maternal parent to 178 inbred lines from DGRP representing a spectrum of natural variation (excluding recessive lethal variants eliminated in the formation of the DGRP). Among several eye phenotypes observed — rough eye, reduced total area, distortion of the oval shape and black lesion spots — we chose total eye area as the phenotype to carry out a GWAS. We quantified eye area in ten male progeny from each hINS$^{C96Y}$ x DGRP cross. We observed a continuously varying distribution of this phenotype, ranging from 13% to 86% of wild type fly eye area (**Figure 1**). ANOVA indicated that nearly 60% of the variance is between genotypes, indicating a large genetic component. Males were chosen for measurement and analysis because they showed a more severe phenotype than females (Park et al., 2013). However, we also measured F1 females for a subset of 38 lines and found a strong correlation between the two sexes from the same cross (r=0.8, **Figure S2**).

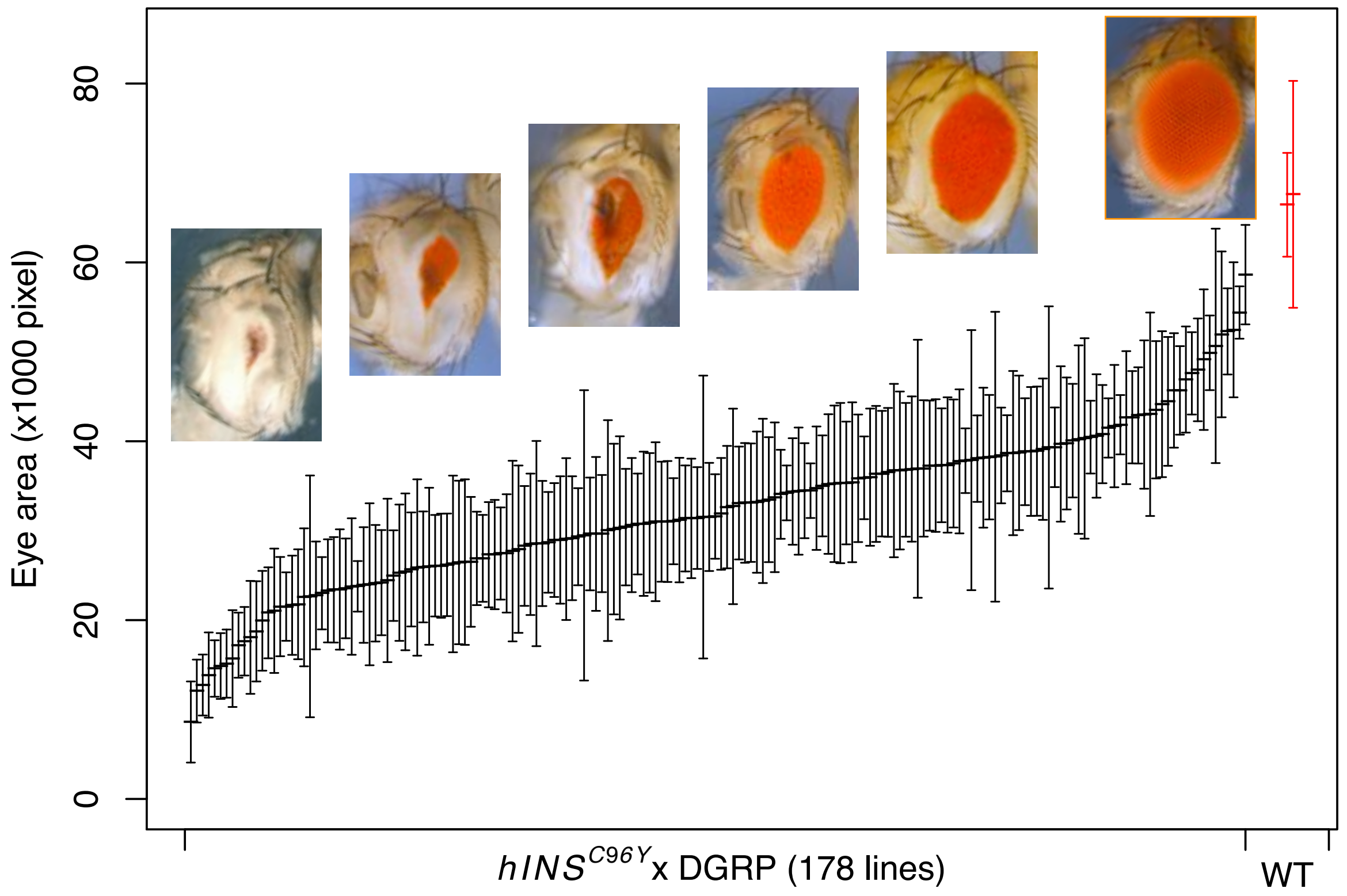

**Figure 1 Distribution of eye area in hINS$^{C96Y}$ x DGRP crosses.** Mean ± 1 s.d., sorted by the mean, is shown for crosses between the transgenic {GMR>>hINS$^{C96Y}$} line to 178 DGRP lines, and two randomly chosen DGRP inbred lines (red). Representative photographs of eyes from across the range of the distribution are shown. The rightmost image is of a non-transgenic wild type fly eye.

The observed variation in eye degeneration is consistent with the hypothesis that it reflects differences in cellular response to the expression of hINS$^{C96Y}$. The severity of the eye degeneration phenotype is not correlated with body size of the same individual, or the mean eye size of the same line; neither is it correlated with GAL4 protein levels in eye imaginal discs (Park et al., 2013). The GWAS described below showed no evidence for association between eye area and SNPs in or surrounding the *glass* (*gl*) locus, the trans-activator of GMR-Gal4, a result consistent with Gal4 protein measurements. Finally, when we expressed mutant hINS in the notum (rather than the eye) and measured the loss of macrochaetae in F1 crosses to 38 DGRP lines for which we also collected eye degeneration data, we observed no correlation between the two traits, indicating that the degeneration phenotypes are not caused by line-specific differences in mutant insulin expression (Park et al., 2013).

***Genome-wide association analysis***

We carried out GWAS on the F1 males from crosses of $hINS^{C96Y}$ and 154 DGRP lines. Population structure is a potential confounding factor in GWAS. Little population structure was found within the full set of DGRP lines (Mackay et al., 2012) and we observed no evidence for population structure in the 154 lines examined using a principal component analysis on 900K autosomal SNPs (obtained by pruning a total of 2 million based on pairwise LD) (**Figure S3**). There was no significant correlation between the $hINS^{C96Y}$ phenotype (line mean) and projection length in the direction of the top five principle components in each DGRP line.

We used mean eye area as a quantitative trait to perform single marker regression for 2.1 million autosomal SNPs. We restricted the analysis to bi-allelic sites for which the minor allele was present in at least four lines (Mackay et al., 2012). Because of the direction of the cross, all F1 males inherited their X-chromosome from the GMR>>$hINS^{C96Y}$ tester line and we expected to observe no association between the phenotype and X-linked SNPs. Quantile-quantile (Q-Q) plots for autosomal and X-chromosomal SNPs revealed that only the former and not the latter showed an excess of small p-values (**Figure 2A,B**). We observed association of a SNP (list SNP) on chromosome 3L with eye area (**Figure 2C**), with a p-value that approached genome-wide significance (raw $P$ = 2.4 x$10^{-8}$, Bonferroni corrected $P$ = 0.0502).

The Bonferroni correction is conservative because whereas it assumes complete independence of SNPs, LD is common between SNPs separated by short distances. To identify additional association beyond the peak on chromosome 3L, we used the threshold $P < 10^{-5}$ suggested by (Mackay et al., 2012) for nominating candidate SNPs. Thirty SNPs, 29 of which were on autosomes and one on the X chromosome, were identified using this criteria (**Table S3**). The 29 autosomal SNPs were distributed in 16 unlinked loci. Two methods were used to assess the false discovery rate (FDR) at this threshold. First, assuming that the FDR for autosomal and X-linked SNPs are the same, we expected 6.5 autosomal SNPs while observing 29, suggesting a FDR of 22%.

Second, we randomized the phenotype relative to the genotype 2,000 times and carried out GWAS on each of the permuted datasets. The resulting number of SNPs passing the threshold in each of the 2,000 trials had a mean of 21 and a median of 19, with the observed number 29 at the 85th percentile. Both methods suggest a modest enrichment of true positives under the $10^{-5}$ threshold.

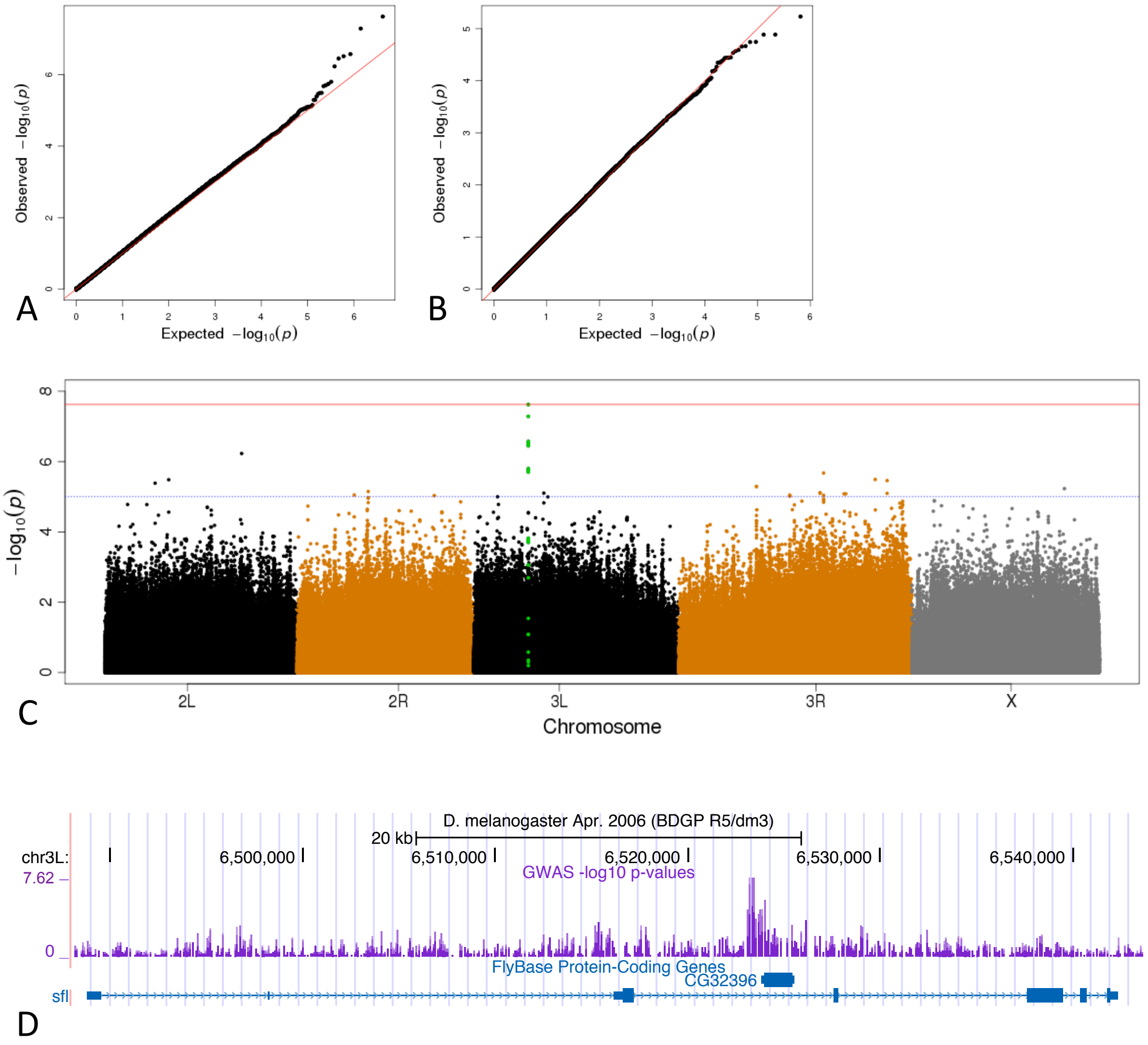

**Figure 2 Genome-wide scan identifies candidate locus associated with the hINS$^{C96Y}$-induced phenotype.** Quantile-Quantile (QQ) plot reveals an excess of small p-values on autosomes (A) but not on the X chromosome (B), which is not variable in the mapping population due to cross design. (C) Manhattan plot shows a strong peak (green) on chromosome 3L. The blue and red horizontal lines indicate the nominating ($P < 10^{-5}$) and the genome-wide threshold (Bonferroni corrected $P < 0.05$), respectively. (D) UCSC browser view of the *sfl* locus containing the association peak. The intron containing the peak also contains a nested gene CG32396.

### ***sulfateless (sfl) modifies eye area phenotype***

The peak on chromosome 3L is confined to the third intron of the gene, *sfl* (**Figure 2D**). This intron also contains a nested gene (CG32396) lying close to the association peak. CG32396 is predicted to encode a protein with a probable tubulin beta-chain. To determine which of the two genes, or possibly both, is responsible for the association signal, we examined the expression pattern of each gene and also used RNAi to knock down gene expression. *sfl* is expressed in the eye-antennal imaginal disc and eye and brain in adults (**Figure S5, S6**). CG32396 has a testis-specific expression pattern in adults, with very low expression in the adult eye (**Figure S5**) and no detectable expression in eye imaginal discs by RT-PCR (**Figure S6, S7**).

RNAi knockdown of either *sfl* or CG32396 in the eye imaginal disc had no measurable effect on eye area. In contrast, RNAi against *sfl*, but not CG32396, significantly decreased mean eye area in the presence of $hINS^{C96Y}$ but not $hINS^{WT}$ (**Figure 3**). These results identify *sfl*, and not CG32396, as the causal gene underlying the association peak.

To test if *sfl* also modifies the $hINS^{C96Y}$-induced phenotype in other tissues, we carried out RNAi knockdown of *sfl* in the developing wing (using a dpp-Gal4 driver) and notum (using an ap-Gal4 driver). In both experiments we observed more severe phenotypes than that caused by $hINS^{C96Y}$ alone (**Figure S8, S9**). However, the interpretation is made complicated by the fact that *sfl* knockdown alone causes mutant phenotypes in these tissues, consistent with previous knowledge (Lin, 2004). At present we cannot distinguish the alternative hypotheses of additive vs. epistatic interactions between *sfl* and $hINS^{C96Y}$.

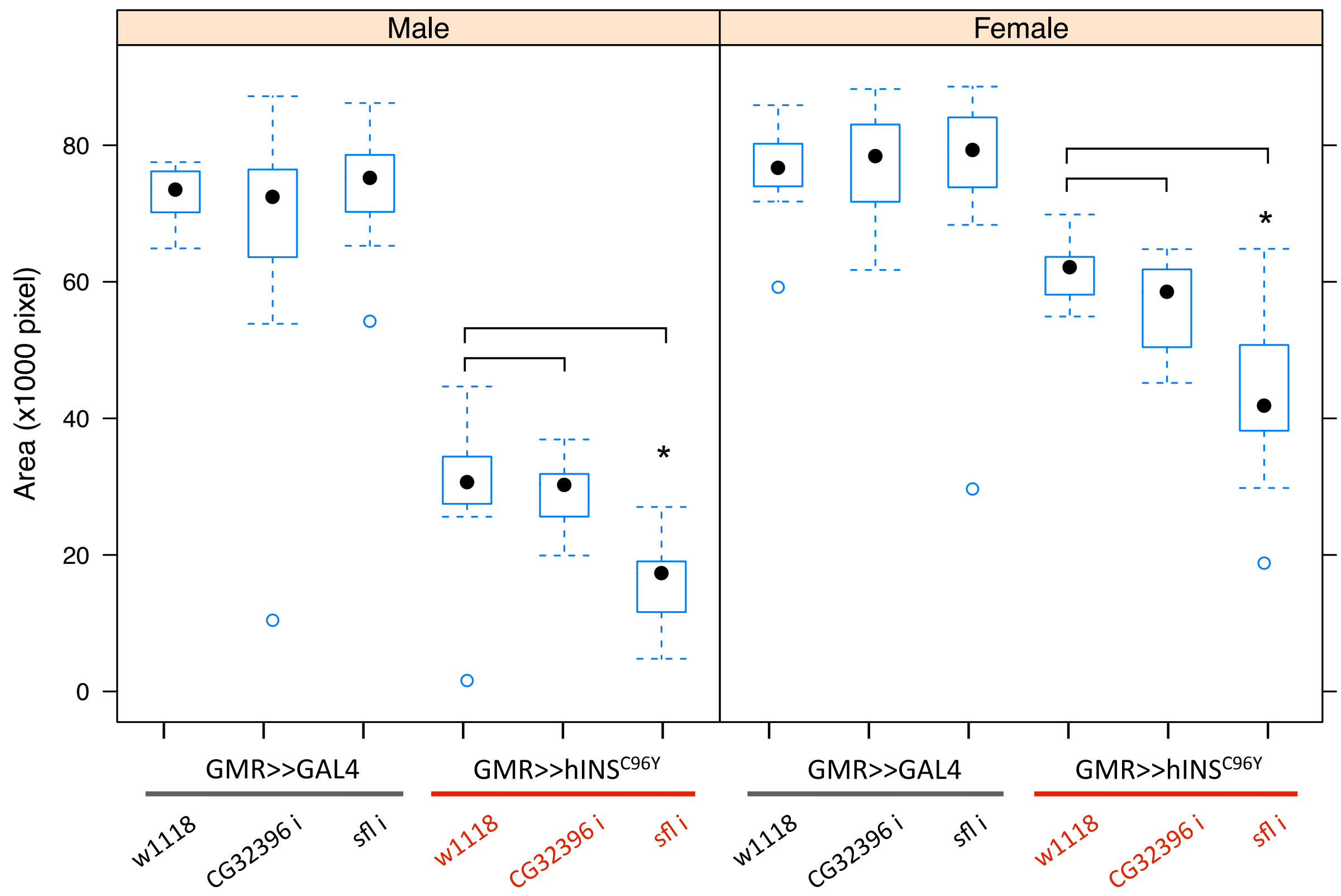

**Figure 3 RNAi knockdown confirms *sfl* and excludes CG32396 as the causal gene.** The effect of knocking down either CG32396 or *sfl* was tested in the absence ({UAS-RNAi} x {GMR-Gal4}) or presence ({UAS-RNAi} x {GMR-Gal4, UAS-hINS$^{C96Y}$}) of hINS$^{C96Y}$. Compared to the control crosses (first and third columns in both sexes), significant difference in mean eye area was observed only with RNAi against *sfl* and only in the presence of hINS$^{C96Y}$ (n=15, asterisks above a box plot indicate significant differences at 0.05 level determined by a student's t-test, with Bonferroni correction for multiple testing). In box plots, the median (black dot), interquartile (box) and 1.5 times the interquartile range (whiskers) are indicated; data points outside the range are represented by circles.

### *Heparan Sulfate Biosynthetic pathway modifies the hINS$^{C96Y}$-induced eye degeneration*

*Sulfateless* encodes a bi-functional enzyme in the HSPG biosynthesis pathway. An important component of the cell surface and extracellular matrix (Kirkpatrick & Selleck, 2007), Heparan sulfate-modified proteins, or proteoglycans (HSPG) are known to regulate signaling during development, influencing the levels and activity of growth factors and morphogens at cell surfaces and in the extracellular matrix (Fujise et al., 2003; Giráldez et al., 2002; Häcker et al., 1997; Kirkpatrick et al., 2004; Nakato et al., 1995). The involvement of HSPGs in the cellular responses to misfolded proteins (proteostasis) has not been previously described.

To further examine the $hINS^{C96Y}$-dependent interaction of *sfl*, we examined RNAi knockdowns and mutants for two additional genes in the HS biosynthetic pathway: *ttv* and *botv*, producing the glycosaminoglycan polymer that is modified by *sfl* (Lin, 2004). Neither of the genes contains significant SNPs in our GWAS (lowest adjusted $P > 0.5$ in both loci, adjusted for multple-testing using either Bonferroni's method or Benjamini & Hochberg method). RNAi knockdown of both genes shows a $hINS^{C96Y}$-dependent effect on eye area in the same direction as *sfl* RNAi (**Figure 4**). In addition, a mutant allele of *botv* also showed a significant dominant enhancement of the eye degeneration phenotype. These results implicate HSPGs in modifying the cellular response to misfolded proteins. Neither of the genes were identified in the GWAS, however, indicating either a lack of disease-affecting variation in these genes in this natural population or the lack of statistical power to detect them.

### *Intronic enhancer and sfl expression*

We re-sequenced a 3 kb region containing the GWAS peak in *sfl* (and the nested gene CG32396) in 19 of the 154 DGRP lines and the transgenic $hINS^{C96Y}$ stock to identify all the variants in this region. We found that the SNP achieving the lowest p-value genome-wide was an 18 bp/4 bp length polymorphism (relative to the *D. simulans* orthologous sequence) (**Figure 5A**). We also found three other insertion/deletion (INDEL) polymorphisms in this region, with sizes ranging from 4 – 30 bp and the minor alleles (deletion in all three cases) being present only once or twice in the sample. In contrast, the 18/4 bp polymorphism is present at 50% frequency in the DGRP sample. In light of the discovery of mislabeled and undiscovered INDELs, we will use the term "Single Feature Polymorphism" (SFP) when referring to variants in the *sfl* locus.

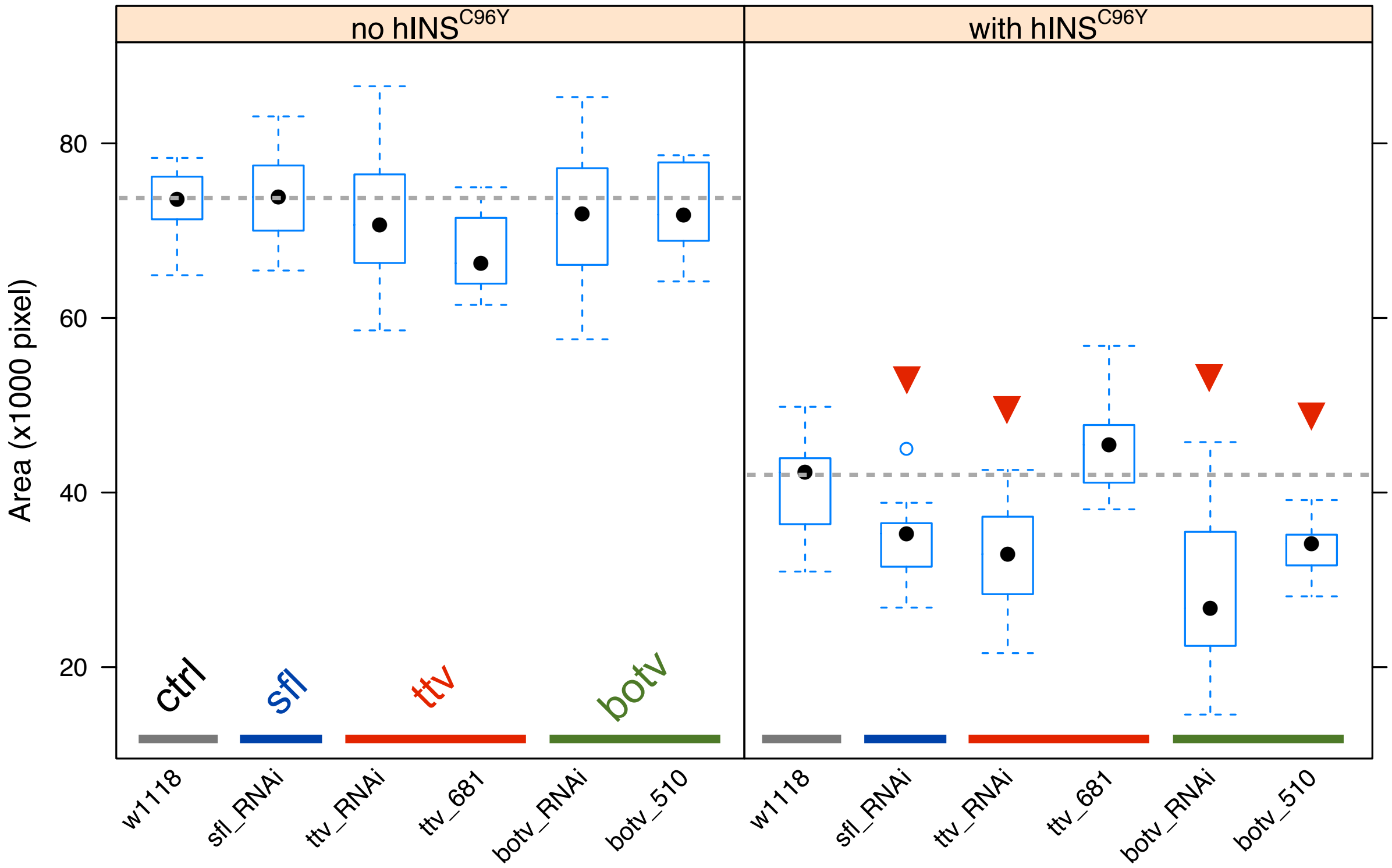

**Figure 4** RNAi and mutant analysis for HSPG biosynthesis pathway genes. The experimental design is the same as in Figure 3. Left panel shows the effect of RNAi or mutant alleles in the absence of hINS$^{C96Y}$ expression; right panel shows the effect when hINS$^{C96Y}$ is expressed in the eye imaginal disc. Mutants were tested in heterozygous states for a dominant interaction with hINS$^{C96Y}$. Fifteen male flies are measured for each group. The statistical significance of differences from the control cross (gray, w1118) was determined by a two-sided student's t test. Those that are significant at 0.05 level after Bonferroni correction are marked with a red arrowhead.

A plot of haplotype structure surrounding the association peak (Haploview v4.2) shows an LD block of 400 bp (block 66 in **Figure 5A**, chr3L:6523119-6523518). There are two major haplotypes, which we name after the 18/4 bp length polymorphism, each represented by two equal-sized groups among the 154 DGRP lines (**Figure 5B**). Because all coding variants in *sfl* lie outside of this 400 bp LD block, we hypothesized that one or more of these intronic SFPs are the causal variant(s) and modify the hINS$^{C96Y}$-induced eye phenotype by altering *sfl* expression.

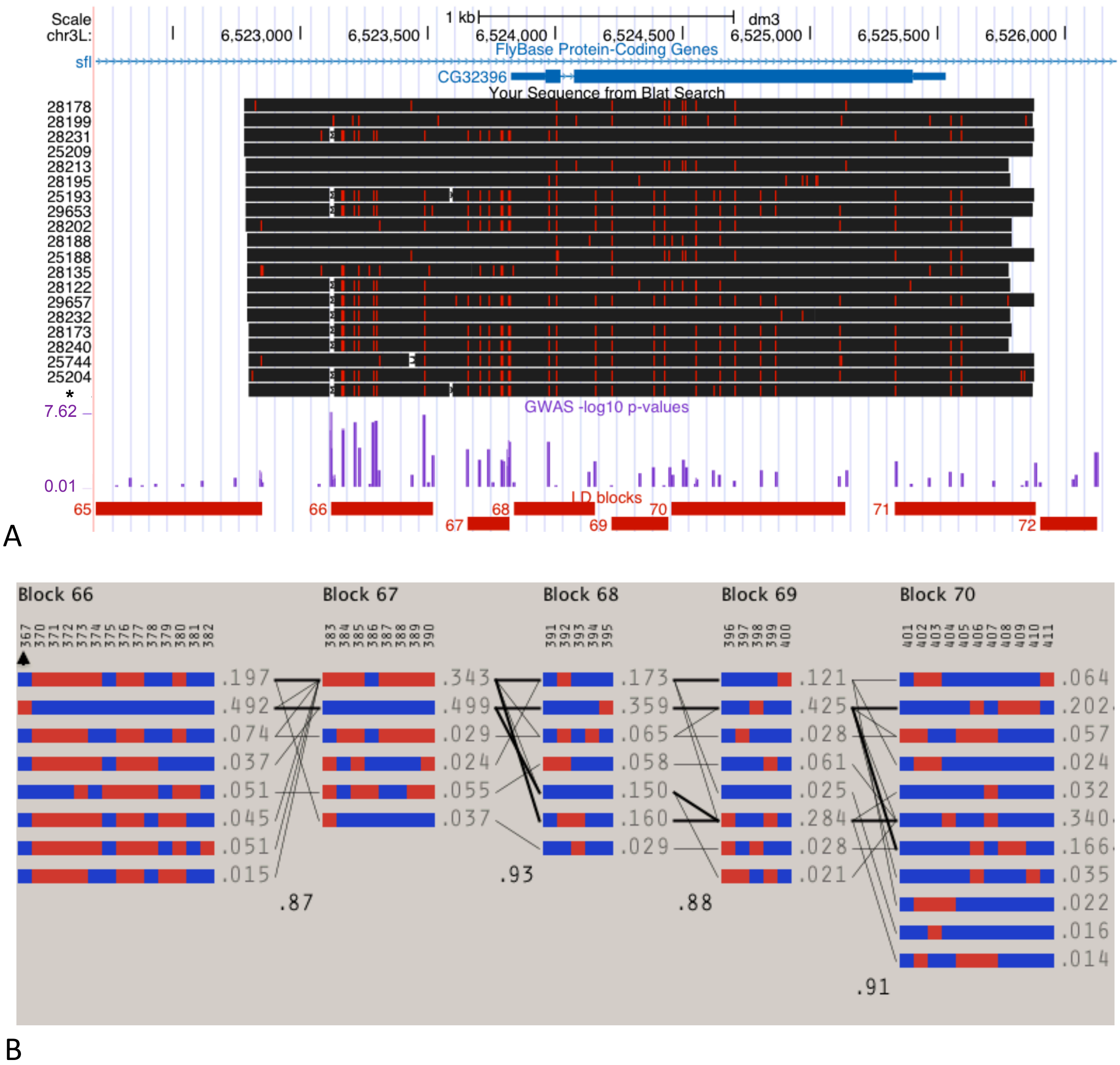

**Figure 5** Sanger re-sequencing of a 3kb region under the peak and the linkage patterns in the region. (A) Alignment of 19 DGRP sequences ordered by their eye degeneration phenotype (mean, most severe on the bottom). The hINS$^{C96Y}$ transgenic line (asterisk) was also sequenced. Red ticks and white spaces indicate SNPs and deletions relative to the reference sequence. No insertions relative to the reference were found. The purple track shows the -log$_{10}$ of GWAS p-values. The bottom track shows the linkage blocks as determined by Haploview (4.02) using the solid spine method with default settings (D' > 0.8). (B) Detailed haplotype block structures. Each numbered column represents a polymorphic site, with the alleles colored as blue or red; each row represents a haplotype with frequency > 0.01. An arrowhead marks the 18/4bp indel polymorphism (see text; 18bp: blue; 4bp: red). Finally, the number between any two blocks represents the multi-allelic D', which quantifies the associations between adjacent blocks.

To test this hypothesis, we crossed randomly selected pairs of 4 bp and 18 bp lines to obtain F1 individuals heterozygous for the two alleles. We then used pyro-sequencing to estimate the relative expression of the two alleles in eye imaginal discs. This method allowed us to measure the ratio of expression of *sfl* associated with each allele in the same animal, thereby controlling for both the trans-environment as well as experimental noise, resulting in highly reproducible results (**Figure S10**). To account for the heterogeneity due to variation in other parts of *sfl* or elsewhere in the genome, we randomly chose six and eight lines carrying the 18 bp or 4 bp allele, respectively, and used them to form 15 crosses (**Figure S1A**). Based on RNAi knock-down of *sfl*, which enhanced the $hINS^{C96Y}$ phenotype, we expected the 4bp allele (associated with more severe phenotypes in the GWAS) to produce less transcript than the 18bp allele.

Allele-specific expression of *sfl* differed in both magnitude and direction among the 15 crosses (**Figure 6**). Seven crosses supported the hypothesis by exhibiting significantly greater expression from the 18 bp allele, with an 18/4 bp ratio ranging from 1.03 - 2.8 (median = 1.15). Two crosses, however, showed slightly greater expression from the 4 bp allele (18 bp/4 bp ratios of 0.94 and 0.96). The remaining six crosses showed no significant differences in expression of the two alleles. The trend towards more expression of the transcript linked to the 18bp allele supports the hypothesis that the 18/4 bp intronic polymorphism (and/or other SFPs in the 400bp LD block) modulates *sfl* expression. However, heterogeneity in allele-specific expression between crosses indicates the presence of additional *cis*-acting expression variants.

### *Search for Additional Association by Conditional Analysis*

In light of the above finding, which suggests additional variants influencing *sfl* expression, we carried out a conditional analysis to identify variants that act independently of the 18bp/4bp SFP. To do so, we tested variants other than the 18/4bp SFP, either within the *sfl* locus or genome-wide, by treating the 18/4bp SFP as a covariate in a linear regression model. After accounting for multiple testing, we observed no significant signals in either case (**Figure S11**). The lack of significance genome-wide may be attributable to the lack of power after correcting for multiple testing. The analysis restricted to the 40kb *sfl* locus reduces the burden of multiple testing by several orders of magnitude, but also fails to identify a significant association. Considering the large range of allele-specific expression differences between the 18bp and 4bp alleles observed in the 15 crosses, the additional *cis*-acting expression variants must either be low frequency alleles or have epistatic properties, two situations this analysis would be underpowered to detect.

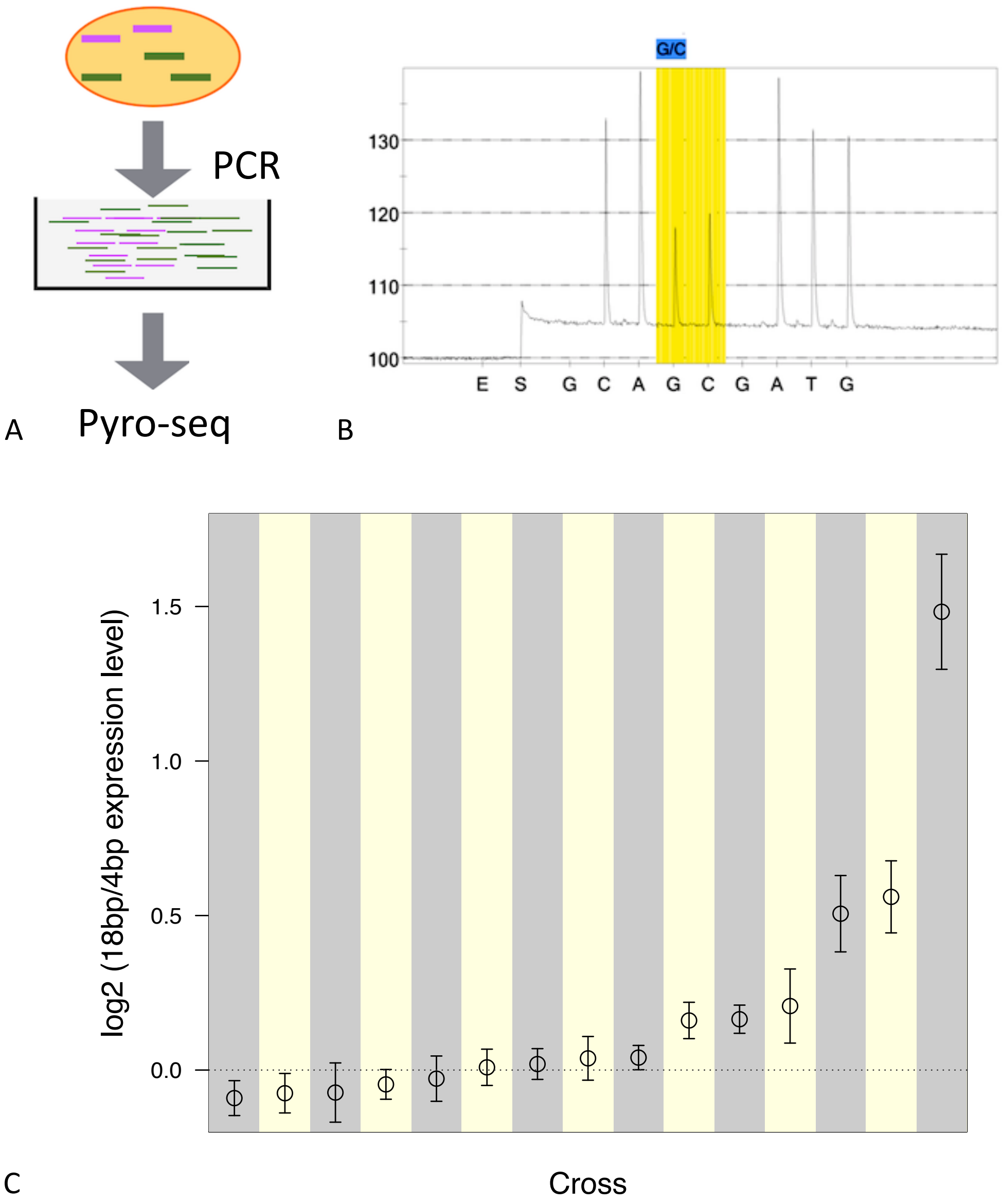

**Figure 6** Pyro-sequencing measure of *sfl* allele-specific transcript ratio in 18bp/4bp heterozygotes. (A) A diagram of the pyro-sequencing approach. Colored lines represent transcripts (mRNA) associated with either the 18bp or the 4bp allele, expressed at different levels. Common primers were used to amplify both transcripts of the gene of interest from the cDNA library made from eye imaginal disc tissues. Pyro-sequencing was carried out on the amplified products. (B) A pyrogram of a heterozygote with the polymorphic site (G/C) that is diagnostic for the 18bp/4bp indel highlighted. The ratio of the two peaks (light intensity, y-axis) are used to calculate the relative ratio of the two alleles. (E: enzyme, S: substrate, A/C/G/T: nucleotides). (C) Log2 transformed ratio of 18bp/4bp allele expression in 15 crosses between randomly paired 18bp and 4bp lines. Estimates of the ratio and 95% confidence intervals are plotted. The dotted line corresponds to equal expression from the two alternative alleles.

## DISCUSSION

### *HSPG function and misfolded protein response*

Our study identified the heparan sulfate biosynthesis pathway (*sfl*, *ttv* and *botv*) as a modifier of eye degeneration induced by expression of a misfolded human proinsulin protein. Although we do not yet know whether this response is to a specific misfolded protein ($hINS^{C96Y}$) or whether it applies to a broader class of misfolded proteins, our discovery now implicates the HSPGs in the regulation of cellular proteostasis.

We propose that genetic variation in HSPG biosynthesis influences the response to misfolded protein through its biological activity in vesicular trafficking of misfolded protein. HSPG are abundant components of cell surfaces and extracellular matrices, and are best understood for their roles in cell signaling and in functioning as co-receptors, both of which are integral to normal development (Häcker et al., 2005; Kirkpatrick & Selleck, 2007). HSPG are also involved in endocytocis (Ren et al., 2009; Stanford et al., 2009) and vesicular trafficking (Nybakken & Perrimon, 2002; Sarrazin et al., 2011), roles that link them to cellular response to misfolded proteins (Higashio & Kohno, 2002; Kim et al., 2009; Kimmig et al., 2012).

It is feasible that HSPGs influence membrane trafficking indirectly, perhaps by regulating signaling events that impinge on trafficking processes. The generation of phosphatidylinositol (3,4,5) triphosphate [PtdIns(3,4,5)$P_3$] by type I phosphoinositide (PI) 3-kinases is affected by a number of growth factors and cytokines, many of which are influenced by HSPGs as accessory molecules. PtdIns(3,4,5)$P_3$ affects a number of trafficking events, including endocytosis and autophagy (Downes et al., 2005).

In a yeast study of the mutant protein folding assistant, protein disulfide isomerase (Pdi1a'), the authors found that more than half of the 130 genes identified as synthetic-lethal were related to vesicle trafficking, while only 10 belonged to the canonical unfolded protein response (UPR) pathway (Kim et al., 2009). In another study, Kimmig *et al* found an enrichment of vesicle-trafficking related genes among those that changed expression significantly after induction of ER-stress (Kimmig et al., 2009). Both studies indicate that a global regulation of vesicle trafficking is important to a cell's response to unfolded or misfolded protein. Activation of UPR has also been shown to affect ER-to-Golgi transport via stimulation of COPII vesicle formation from the ER (Higashio & Kohno, 2002). We therefore propose that either natural variation or genetic perturbation of HS biosynthesis influences the global regulation of vesicle trafficking, which in turn affects cells' ability to process an excess of unfolded or misfolded protein. Prolonged ER-stress may then lead to apoptosis.

### *Genetic Architecture of the* **hINS$^{C96Y}$***-induced eye degeneration phenotypes*

Phenotypic heterogeneity that is dependent on the genetic background is a common phenomenon, and in humans, imposes a significant challenge in both diagnosis and treatment. Our fly model provides a tractable system for studying the genetic and molecular basis for such phenotypic heterogeneity, but with limitations imposed by the sample size of the study. *sfl* is the only significant QTL identified under a highly stringent threshold. A less stringent cutoff of $P < 10^{-5}$ identifies 29 SNPs in 16 loci; permutation test, however, finds this number at 85% percentile among 2,000 randomly shuffled datasets, suggesting that the false discovery rate may be high. A simple calculation for a t-test based statistic at $P = 0.05$ level with Bonferroni's correction for multiple testing indicates that we will have 66% power to identify a variant at 50% population frequency, with an effect-size of 1 (measured as the shift in phenotypic mean in units of standard deviation of the trait, see **Table S4**). This example was chosen to match the estimates for the 18bp/4bp indel polymorphism in the *sfl* intron in the sample of 154 crosses. Any variant with a smaller effect-size and/or lower frequency than the 18bp/4bp polymorphism would have been missed in this study with a high probability.

The genetic architecture for the hINS$^{C96Y}$-induced eye phenotype must involve many loci in addition to *sfl*, as is evident from the continuous distribution of the between-line phenotypes in crosses between the DGRP lines and GMR>>hINS$^{C96Y}$ tester strain (**Figure 1**). We applied a mixed-linear model to the mean phenotype for the 154 crosses to estimate the total contribution of common variants to the phenotypic variance (Yang et al., 2010). This method does not identify individual SNP, and thus does not suffer from the multiple-testing burden as in GWAS. Applying this method (Yang et al., 2011), 52% (standard error = 28%) of the variance between crosses (i.e. genotypes) can be attributed to common, autosomal variants with minor allele frequencies greater than 5%. Although this estimate has a large standard error, resulting from the limited number of lines, it nevertheless indicates contributions by many additive factors beyond *sfl*, most of which likely have small effect sizes or low population frequencies, making them not detectable in our study.

An additional layer of genetic complexity was revealed when we tested the effects of the intronic SFPs on *sfl* expression. Among the 15 crosses where we randomly paired an 18bp allele line with a 4bp allele line from DGRP, seven showed greater expression from the 18bp-linked allele than the 4bp-linked allele; six showed no significant differences while two showed small, yet significant differences in the opposite direction. While the general trend is as expected (**Figure 6**), the phenotypic heterogeneity suggested that additional variation is influencing *sfl* expression. The variants most likely lie within the *sfl* locus because the experiment specifically measures allele-specific expression in heterozygotes, which means any trans-variant will have to act in an allele-specific manner. A conditional analysis using the 18bp/4bp polymorphism as a covariate failed, however, to identify secondary SFPs in the *sfl* locus, most likely due to lack of power.

### *Relationship to common, complex diseases*

While our fly model is of a monogenic form of diabetes, it exhibits a complex genetic architecture when placed on a diverse set of genetic backgrounds. Here we would like to argue that fly models of monogenetic disease are suitable subjects for the genetic dissection of common disorders in human.

One role of the Mendelian mutation is to sensitize the fly to allow phenotypic effects of background genetic modifiers to become visible. Although common disorders are normally considered as lacking a major mutation, a careful consideration suggests that this view is inaccurate. What common disorders lack are large-effect mutations shared by a substantial proportion of the affected individuals. For many diseases, perturbation may be required to boost the expressivity of additive genetic variation that would otherwise be cryptic, i.e., below a disease-causing threshold. Such a perturbation could be genetic, such as driver mutations in cancer, but could also be environmental, such as diet and life-style changes in the case of cardiovascular disease and type 2 diabetes. Consistent with this view, it has been proposed that recent genome evolution and rapid environmental as well as cultural changes in human history have decanalizing effects on physiology, which release cryptic genetic variation and underlie the rising incidence of common human disorders (Gibson, 2009).

A genetic screen for naturally occurring modifiers in a sensitized background, such as the one we employed here, should apply equally well in the study of Mendelian or complex disease. Were this not the case, two different classes of genetic modifiers would have to be posited. An intriguing question, which we found little empirical evidence for or against, could be addressed in the fly by constructing a series of sensitized backgrounds utilizing different disease-causing mutant hINS alleles of varying effect on disease (*e.g.,* neonatal diabetes *vs.* maturity-onset diabetes of the young Støy et al., 2007), and comparing the composition of naturally occurring modifiers.

### *Advantages of a fly model of complex disease*

A primary mutation can manifest itself in different ways and with tissue-specific effects (Mefford et al., 2008), possibly a consequence of its interdependence with the individual's genetic background. The binary Gal4-UAS system enables the creation of a series of models using the same disease mechanism, but directed to different tissues with high tissue-specificity. The ability to construct and study multiple related models in parallel can provide insight into the basis of disease heterogeneity. In the accompanying paper we show, for example, that the developing eye and notum have different sets of genetic background modifiers of mutant hINS-dependent disease (Park et al., 2013). Sex-specific difference in disease risk and severity are also readily modeled in the fly. In both the fly and mouse model of $hINS^{C96Y}$-induced disease, males consistently show more severe disease phenotypes (Wang et al., 1999, Park et al., 2013).

*Drosophila* models of human disease provide a useful alternative to the study of a complex disease in patient populations. First, many models of human disease have been established in the fly, most notably neurodegeneration and cancer (Bilen & Bonini, 2005; Gonzalez, 2013). We predict that natural variation will influence the severity of disease phenotypes in all of them. Second, many models of disease can be created by over-expression of a mutant allele, which makes them suitable for F1 screens between a tester stock and inbred population collections, such as we employed here. Our study shows that an impressive amount of genetic variation for disease severity is dominant or semi-dominant; this outcrossing design also avoids unwanted effects of inbreeding on traits and better mimics the natural heterozygosity of low frequency variants. Third, this experimental design facilitates repeated measurement of a disease phenotype, thereby increasing the power to detect a causal association (Mackay et al., 2009). Fourth, LD is low in *D. melanogaster* and SNP is 20-40X more abundant than in human. Finally, both forward and reverse genetics can be applied to investigate the biology and pathway genetics of candidate variants. For all these reasons we believe fly models will prove useful in understanding the genetic architecture of complex human disease.

## ACKNOWLEDGMENTS

This work was funded by grants from the National Institute of Diabetes and Digestive and Kidney Diseases (R01 DK013914 and P30 DK020595), the National Institute of General Medical Sciences (GM081892), the Chicago Biomedical Consortium with support from the Searle Funds at The Chicago Community Trust, and a gift from the Kovler Family Foundation. SBS is supported by GM054832 and PJW is supported by NSF MCB-1021398. We thank Dan Nicolae, Xiang Zhou for technical help and advice on GWAS. We thank Jian Yang and Peter Visscher for help with the mixed linear model analysis. We thank Joseph Coolon in the Wittkopp lab for designing the pyro-sequencing assays, and Ellen Pederson at the DNA sequencing center at U of Michigan for providing technical assistance.

## SUPPLEMENTARY TABLES

**Table S1** DRRP lines used in this study

| Line | Stock |
|---|---|
| RAL-101 | 28138 |
| RAL-105 | 28139 |
| RAL-109 | 28140 |
| RAL-129 | 28141 |
| RAL-136 | 28142 |
| RAL-138 | 28143 |
| RAL-142 | 28144 |
| RAL-149 | 28145 |
| RAL-153 | 28146 |
| RAL-158 | 28147 |
| RAL-161 | 28148 |
| RAL-176 | 28149 |
| RAL-177 | 28150 |
| RAL-195 | 28153 |
| RAL-208 | 25174 |
| RAL-21 | 28122 |
| RAL-217 | 28154 |
| RAL-227 | 28156 |
| RAL-228 | 28157 |
| RAL-229 | 29653 |
| RAL-233 | 28159 |
| RAL-235 | 28275 |
| RAL-237 | 28160 |
| RAL-239 | 28161 |
| RAL-256 | 28162 |
| RAL-26 | 28123 |
| RAL-28 | 28124 |
| RAL-280 | 28164 |
| RAL-287 | 28165 |
| RAL-301 | 25175 |
| RAL-303 | 25176 |
| RAL-304 | 25177 |
| RAL-307 | 25179 |
| RAL-309 | 28166 |
| RAL-310 | 28276 |
| RAL-313 | 25180 |
| RAL-315 | 25181 |
| RAL-318 | 28168 |
| RAL-320 | 29654 |
| RAL-321 | 29655 |
| RAL-324 | 25182 |
| RAL-325 | 28170 |
| RAL-332 | 28171 |
| RAL-335 | 25183 |
| RAL-338 | 28173 |
| RAL-350 | 28176 |
| RAL-352 | 28177 |
| RAL-356 | 28178 |
| RAL-357 | 25184 |
| RAL-358 | 25185 |
| RAL-359 | 28179 |
| RAL-360 | 25186 |
| RAL-362 | 25187 |
| RAL-365 | 25445 |
| RAL-367 | 28181 |
| RAL-371 | 28183 |
| RAL-373 | 28184 |
| RAL-374 | 28185 |
| RAL-375 | 25188 |
| RAL-377 | 28186 |
| RAL-379 | 25189 |
| RAL-38 | 28125 |
| RAL-380 | 25190 |
| RAL-381 | 28188 |
| RAL-383 | 28190 |
| RAL-386 | 28192 |
| RAL-391 | 25191 |
| RAL-399 | 25192 |
| RAL-40 | 29651 |
| RAL-405 | 29656 |
| RAL-406 | 29657 |
| RAL-409 | 28278 |
| RAL-41 | 28126 |
| RAL-42 | 28127 |
| RAL-426 | 28196 |
| RAL-427 | 25193 |
| RAL-437 | 25194 |
| RAL-439 | 29658 |
| RAL-440 | 28197 |
| RAL-441 | 28198 |
| RAL-443 | 28199 |
| RAL-45 | 28128 |
| RAL-461 | 28200 |
| RAL-486 | 25195 |
| RAL-491 | 28202 |
| RAL-492 | 28203 |
| RAL-502 | 28204 |
| RAL-508 | 28205 |
| RAL-509 | 28206 |
| RAL-513 | 29659 |
| RAL-517 | 25197 |
| RAL-531 | 28207 |
| RAL-535 | 28208 |
| RAL-555 | 25198 |
| RAL-563 | 28211 |
| RAL-57 | 29652 |
| RAL-589 | 28213 |
| RAL-59 | 28129 |
| RAL-595 | 28215 |
| RAL-639 | 25199 |
| RAL-642 | 28216 |
| RAL-646 | 28217 |
| RAL-69 | 28130 |
| RAL-703 | 28218 |
| RAL-705 | 25744 |
| RAL-707 | 25200 |
| RAL-712 | 25201 |
| RAL-714 | 25745 |
| RAL-716 | 28219 |
| RAL-721 | 28220 |
| RAL-73 | 28131 |
| RAL-730 | 25202 |
| RAL-732 | 25203 |
| RAL-737 | 28222 |
| RAL-738 | 28223 |
| RAL-75 | 28132 |
| RAL-761 | 28227 |
| RAL-765 | 25204 |
| RAL-774 | 25205 |
| RAL-776 | 28229 |
| RAL-783 | 28230 |
| RAL-786 | 25206 |
| RAL-787 | 28231 |
| RAL-790 | 28232 |
| RAL-796 | 28233 |
| RAL-799 | 25207 |
| RAL-801 | 28234 |
| RAL-802 | 28235 |
| RAL-805 | 28237 |
| RAL-808 | 28238 |
| RAL-810 | 28239 |
| RAL-812 | 28240 |
| RAL-818 | 28241 |
| RAL-820 | 25208 |
| RAL-822 | 28244 |
| RAL-83 | 28134 |
| RAL-837 | 28246 |
| RAL-85 | 28274 |
| RAL-852 | 25209 |
| RAL-855 | 28251 |
| RAL-857 | 28252 |
| RAL-859 | 25210 |
| RAL-861 | 28253 |
| RAL-879 | 28254 |
| RAL-88 | 28135 |
| RAL-882 | 28255 |
| RAL-884 | 28256 |
| RAL-887 | 28279 |
| RAL-890 | 28257 |
| RAL-894 | 28259 |
| RAL-897 | 28260 |
| RAL-908 | 28263 |
| RAL-91 | 28136 |
| RAL-911 | 28264 |

**Table S2** Sequence primers used in this study

| Name | Sequence (5'->3') |
|---|---|
| | **qRT-PCR** |
| sfl_F1 | TCGATACGGGCGTGTTTAATGGAC |
| sfl_R1 | TTGATAATGGGTGCGGGATGCG |
| CG32396_F1 | AGCGGAGATTGGGTCGAAATGAG |
| CG32396_R1 | CATGTGAAATCACGTGCCAGAAAG |
| kl-3F1 | ATGGCAAACGTAGACCCACCTC |
| kl-3R1 | GTACCGGCGGACGATTCTTTAG |
| Pp1-Y2F1 | TTTGTTGTCACGGCGGTCTCAG |
| Pp1-Y2R1 | ACGTCACATGGTCGGGCTAATTG |
| RP49-F1 | CGGATCGATATGCTAAGCTGT |
| RP49-R1 | GCGCTTGTTCGATCCGTA |
| | **Pyro-seq** |
| 1336F1 | CGGGCGGCAATCAACATAA |
| 1336R1 | CGGTCACGGAGCTACCAAATT |
| 1336S1 | CTCATTAAGCAGCCG |
| 2789F1 | GACTGCGACCAGATGATGTGAG |
| 2789R1 | CTTCCCTCGTGCCATGATGATA |
| 2789S1 | TTCCCGAGAATCCCA |
| 2854F1 | CGGGAAAATACTATCATCATGGC |
| 2854R1 | GTGCGAAAACCAGTTGAACTC |
| 2854S1 | TCCTGAACGTTCTGC |
| 1885F1 | TAATGGACTTATTCAACGCGACAC |
| 1885R1 | TGTGTTTGCCACCAGAGTTG |
| 1885S1 | CGGCAGTTGATAATGG |

**Table S3** Annotation of SNPs below $10^{-5}$ p-value threshold

| Chr | Position | Gene Symbol | Site Class | MAF | P |
|---|---|---|---|---|---|
| 2L | 6018257 | H2.0 | intronic | 0.15 | 4.11E-06 |
| 2L | 7639113 | CG13792 | intergenic | 0.42 | 3.28E-06 |
| 2L | 7639113 | CG6739 | intergenic | 0.42 | 3.28E-06 |
| 2L | 16378839 | CG5888 | intergenic | 0.11 | 5.89E-07 |
| 2L | 16378839 | jhamt | intergenic | 0.11 | 5.89E-07 |
| 2R | 6830823 | Spn47C | intergenic | 0.05 | 8.91E-06 |
| 2R | 6830823 | luna | intergenic | 0.05 | 8.91E-06 |
| 2R | 8514952 | CG17760 | intronic | 0.08 | 7.11E-06 |
| 2R | 16411003 | CG13422 | intergenic | 0.17 | 9.29E-06 |
| 2R | 16411003 | CG13426 | intergenic | 0.17 | 9.29E-06 |
| 3L | 6523119 | sfl | intronic | 0.47 | 2.38E-08 |
| 3L | 6523164 | sfl | intronic | 0.43 | 1.98E-06 |
| 3L | 6523166 | sfl | intronic | 0.43 | 1.82E-06 |
| 3L | 6523167 | sfl | intronic | 0.42 | 1.59E-06 |
| 3L | 6523212 | sfl | intronic | 0.49 | 3.56E-07 |
| 3L | 6523285 | sfl | intronic | 0.48 | 3.08E-07 |
| 3L | 6523298 | sfl | intronic | 0.49 | 2.67E-07 |
| 3L | 6523484 | sfl | intronic | 0.42 | 5.18E-08 |
| 3L | 8372543 | ImpE1 | intronic | 0.21 | 7.87E-06 |
| 3R | 9282003 | CG14372 | intronic | 0.05 | 5.12E-06 |
| 3R | 9282011 | CG14372 | intronic | 0.05 | 5.12E-06 |
| 3R | 13265256 | CG5873 | intronic | 0.03 | 8.95E-06 |
| 3R | 13265256 | CG14331 | intronic | 0.03 | 8.95E-06 |
| 3R | 13265265 | CG5873 | intronic | 0.03 | 9.52E-06 |
| 3R | 13265265 | CG14331 | intronic | 0.03 | 9.52E-06 |
| 3R | 13265268 | CG5873 | intronic | 0.03 | 9.52E-06 |
| 3R | 13265268 | CG14331 | intronic | 0.03 | 9.52E-06 |
| 3R | 16891400 | AnnIX | intronic | 0.45 | 7.66E-06 |
| 3R | 16891456 | AnnIX | intronic | 0.45 | 7.97E-06 |
| 3R | 17323042 | C15 | intergenic | 0.43 | 9.08E-06 |
| 3R | 17323042 | CG7922 | intergenic | 0.43 | 9.08E-06 |
| 3R | 17323100 | C15 | intergenic | 0.32 | 2.11E-06 |
| 3R | 17323100 | CG7922 | intergenic | 0.32 | 2.11E-06 |
| 3R | 19762489 | Gdh | intronic | 0.39 | 8.26E-06 |
| 3R | 19968993 | kal-1 | intronic | 0.35 | 8.25E-06 |
| 3R | 23486244 | CG18437 | intergenic | 0.49 | 3.24E-06 |
| 3R | 23486244 | Mlc1 | intergenic | 0.49 | 3.24E-06 |
| 3R | 24918157 | CG11873 | intronic | 0.49 | 3.48E-06 |
| 3R | 24920070 | CG11873 | intronic | 0.41 | 8.00E-06 |
| X | 18241719 | CG6123 | intronic | 0.03 | 5.88E-06 |

**Table S4** Power calculation for GWAS with 154 lines

| | Effect Size* | | | | |
|---|---|---|---|---|---|
| MAF | **0.75** | **1** | **1.25** | **1.5** | **2** |
| **0.01** | 0.00 | 0.00 | 0.00 | 0.00 | 0.00 |
| **0.05** | 0.00 | 0.00 | 0.01 | 0.03 | 0.26 |
| **0.1** | 0.00 | 0.02 | 0.12 | 0.39 | 0.94 |
| **0.2** | 0.02 | 0.19 | 0.63 | 0.94 | 1.00 |
| **0.3** | 0.06 | 0.45 | 0.90 | 1.00 | 1.00 |
| **0.4** | 0.11 | 0.60 | 0.96 | 1.00 | 1.00 |
| **0.5** | 0.13 | 0.66 | 0.97 | 1.00 | 1.00 |

* Effect size is measured as the shift in the phenotype mean in units of s.d. for the trait

The calculation is done using the t-distribution. The R-code is attached below:

```
myPower.t <- function(effect.size=1,alpha=0.05,m,n){
  ## Power for GWAS t test
  ## calculate power for a t test comparing two populations with equal variance but unequal
sample sizes
  ## m, n: sample size of each allele class, not to be confused with m above
  df = m+n-2
  A = 1/sqrt(1/m+1/n) ## factor for calculating t statistics
  T = qt(1-alpha/2,m+n-2)
  T1 <- T-effect.size*A
  beta <- pt(T1,m+n-2)
  return(1-beta)
}
## plot power of GWAS t test ##
alpha1=.05/1.37e6
power <- NULL
effect.size <- c(0.75,1,1.25,1.5,2)
freq <- c(0.01,0.05,0.1,0.2,0.3,0.4,0.5)
N = 154 # size of GWAS mapping population
for(p in freq){
  m = as.integer(N*p)
  n = N-m
  power <- rbind(power, sapply(effect.size,function(x) myPower.t(x,alpha1,m,n)))
}
dimnames(power) <- list("freq"=freq,"effect.size"=effect.size)
```

## SUPPLEMENTARY FIGURES

A

| | 18bp | 28190 | 28141 | 28178 | 28144 | 28135 | 28171 |
|---|---|---|---|---|---|---|---|
| 4bp | | A | B | C | D | E | F |
| 28240 | 1 | **A1** | **1B** | | | | |
| 28231 | 2 | | **B2** | **2C** | | | |
| 28138 | 3 | | | **C3** | | | **3F** |
| 25204 | 4 | | | | **D4** | **4E** | |
| 28211 | 5 | **5A** | | | | **E5** | **5F** |
| 28227 | 6 | | | | **6D** | | **F6** |
| 28139 | 7 | | | | | | **F7** |
| 28122 | 8 | | | **C8** | | | |

B

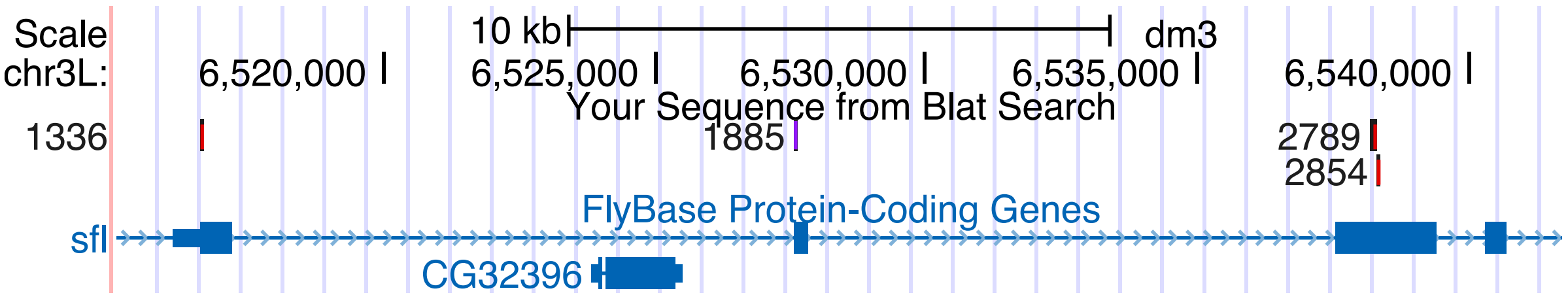

**Figure S1** Pyro-sequencing cross and assay design. (A) Cross design for pyro-sequencing. Six 18bp and eight 4bp lines were randomly chosen from the 154 DGRP lines used in GWAS. The Bloomington center stock number is listed. In each cell, the order of the letter/number indicate the direction of the cross. For example, A1 indicates that males of #28240 was crossed to virgin females of #28190. (B) pyro-sequencing assays. Four SNPs were selected within the transcribed regions so as to distinguish alleles associated with the 18/4 bp indel polymorphism.

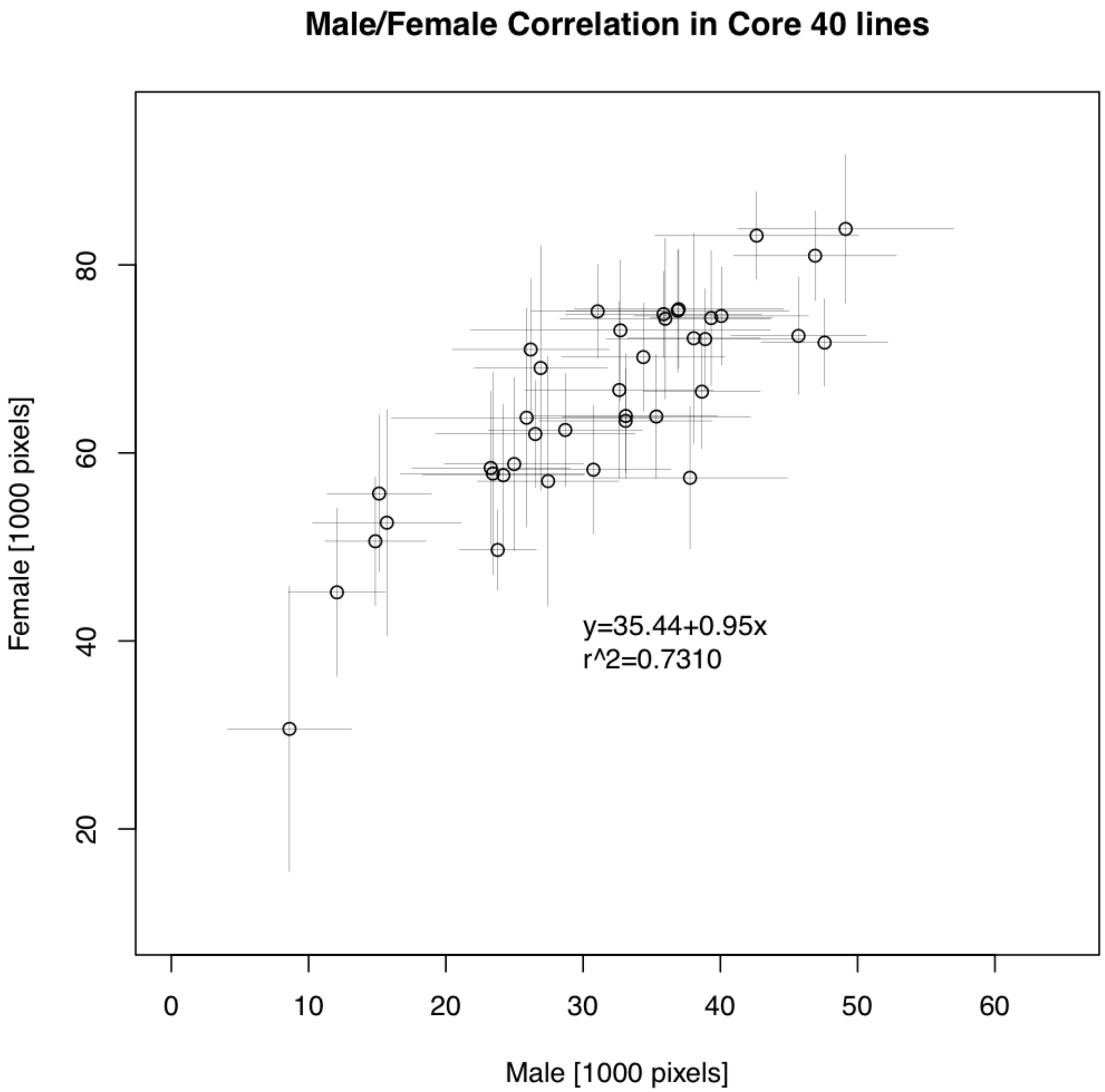

**Figure S2** Correlations of eye area between F1 males and females within the same cross. Mean ± 1 s.d. are plotted for a subset of 38 lines. The least square linear fit is indicated.

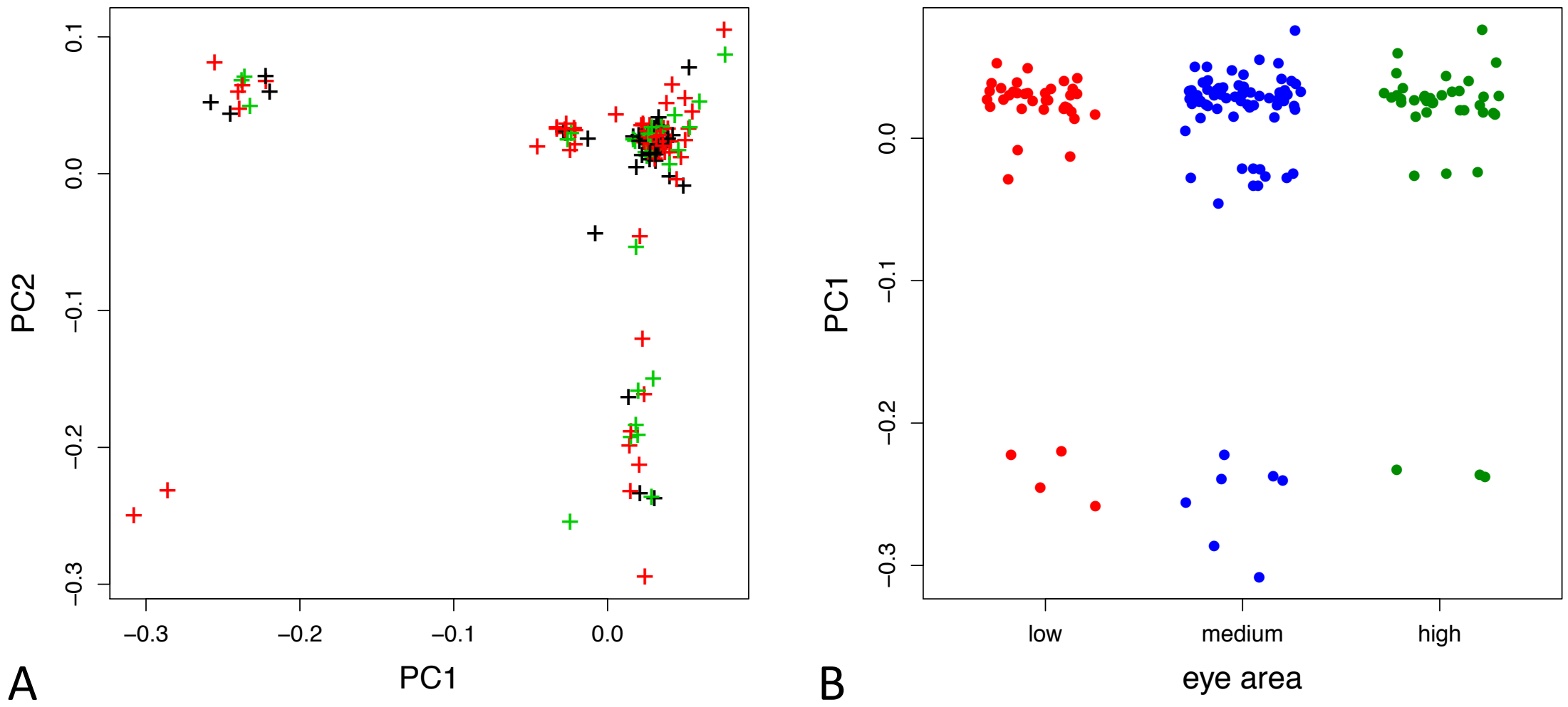

**Figure S3** Population structure assessed through principal component analysis (PCA) using 900K autosomal SNPs after LD pruning. (A) 154 DGRP inbred lines projected onto the plane spanned by the first two principal components (PC1, PC2). The points are colored according to the phenotype severity in the $hINS^{C96Y}$ crosses (red: severe, or first 25%; blue: intermediate, 25%-75%; green: mild, 75%-100%, percentiles in eye area distribution from small to large). (B) projection onto PC1 grouped by their phenotype severity showed no correlation between the two.

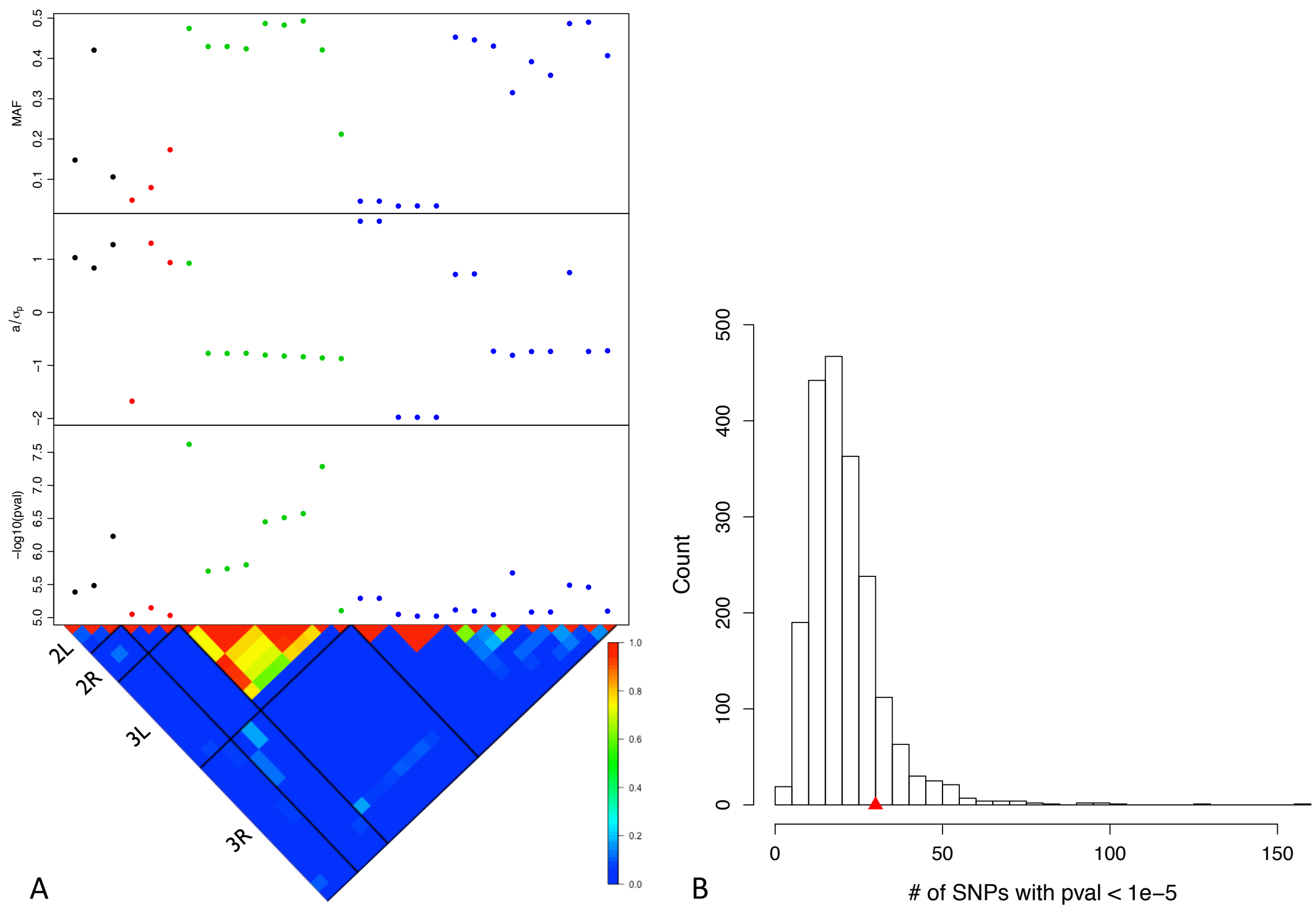

**Figure S4** All SNPs with p-values below $10^{-5}$ from GWAS. (A) Composite plot: from top to bottom are minor allele frequencies, effect sizes, -log10 (p-values) and the bottom triangle represent the linkage map between these SNPs grouped by chromosomes. (B) Distribution of the number of SNPs with a p-value < $10^{-5}$ in 2000 permutations, where the line identities in each trial was shuffled randomly and the same association analysis was applied. Red triangle indicate the observed number of SNPs in the real data, which is at 85th percentile among the 2000 permutations.

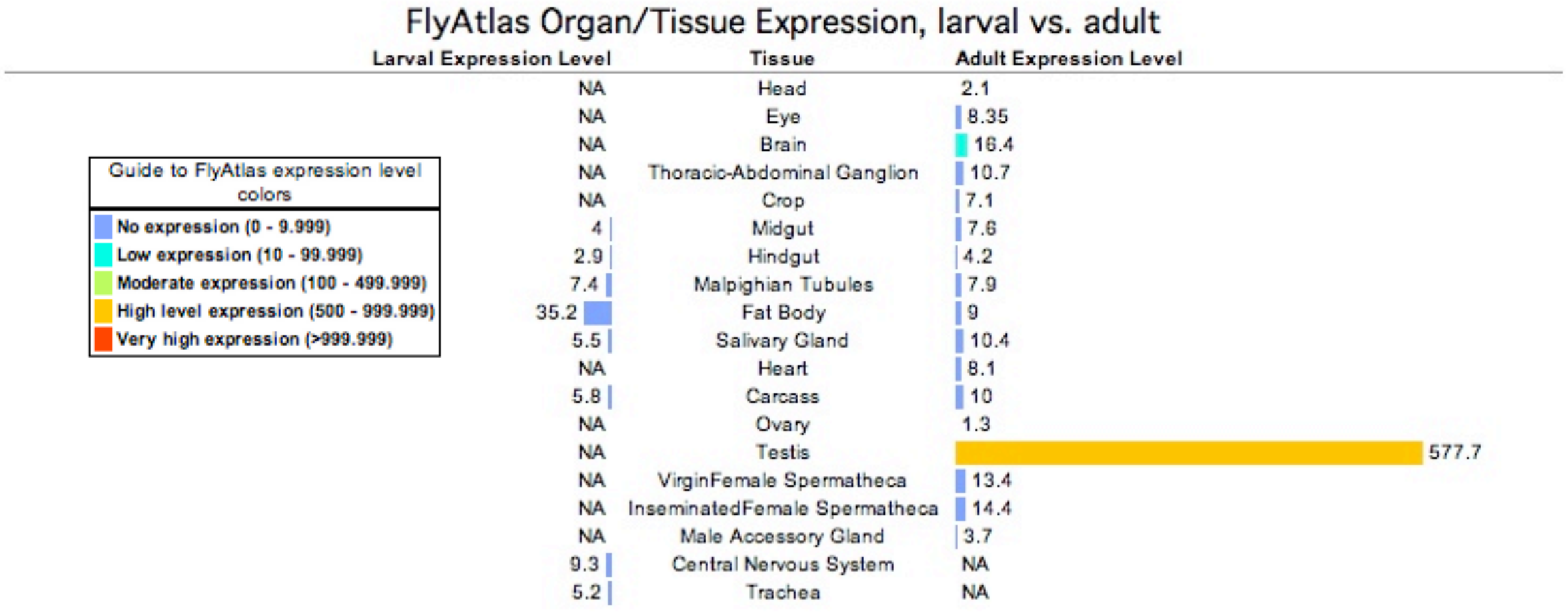

A

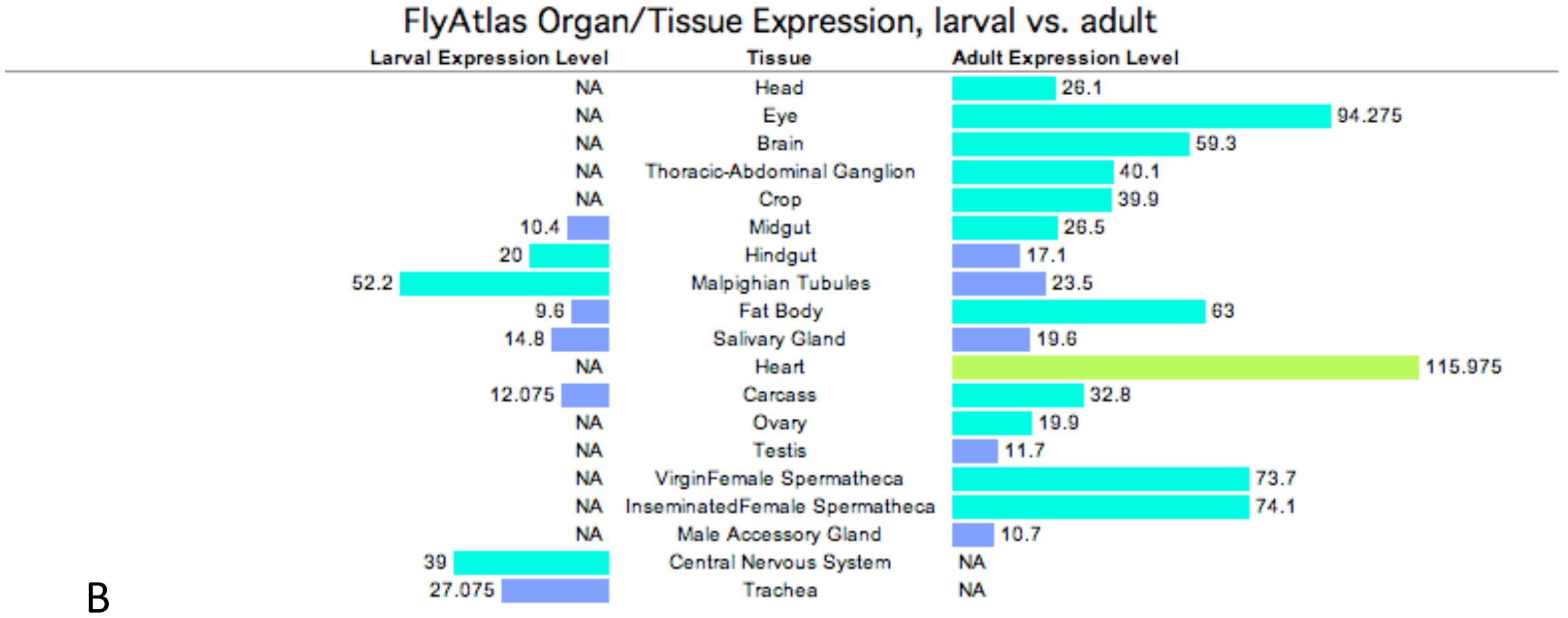

B

**Figure S5** FlyAtlas expression report for CG32396 and sfl. (A) CG32396 (B) sfl. Figure obtained through FlyBase.

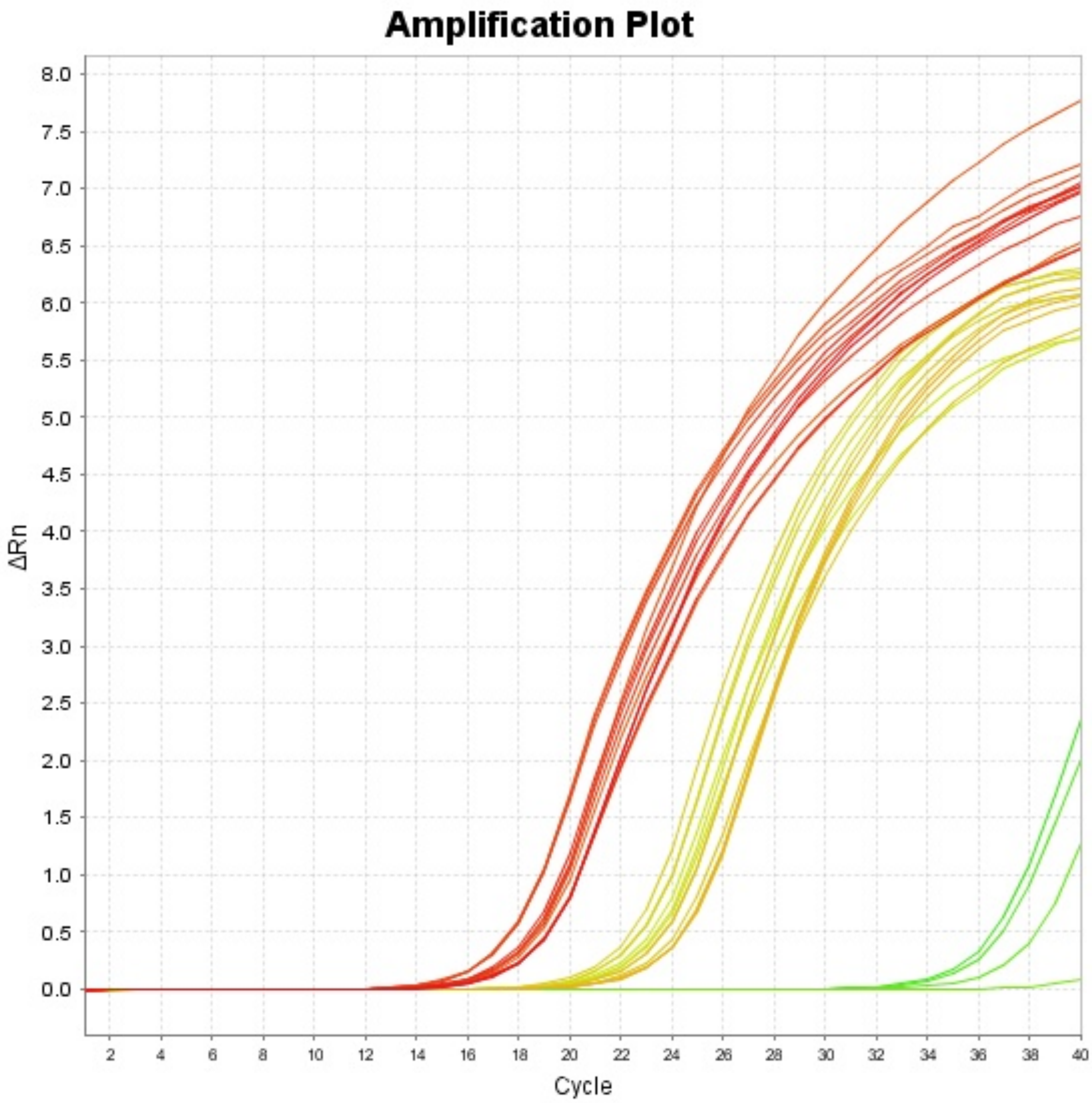

**Figure S6** qRT-PCR quantification of mRNA levels for CG32396 and *sfl* in eye imaginal disc samples.  Two inbred lines from DGRP were randomly chosen and eye imaginal disc samples were prepared from either 6 male or 6 female larvae, resulting in 4 biological samples. qRT-PCR were performed for each sample and three genes (RP49 -- red curve, *sfl* -- yellow, and CG32396 -- green). Shown is the amplification plot: x-axis -- cycle number; y-axis -- base-line corrected relative fluorescence intensity proportional to the amount of amplicons. Both RP49 and *sfl* were detected starting in the 18-20th cycle, while amplification didn't happen for CG32396 until after 32 cycle. In addition, multiple melting points were detected for CG32396 assays, but not in the other two genes.

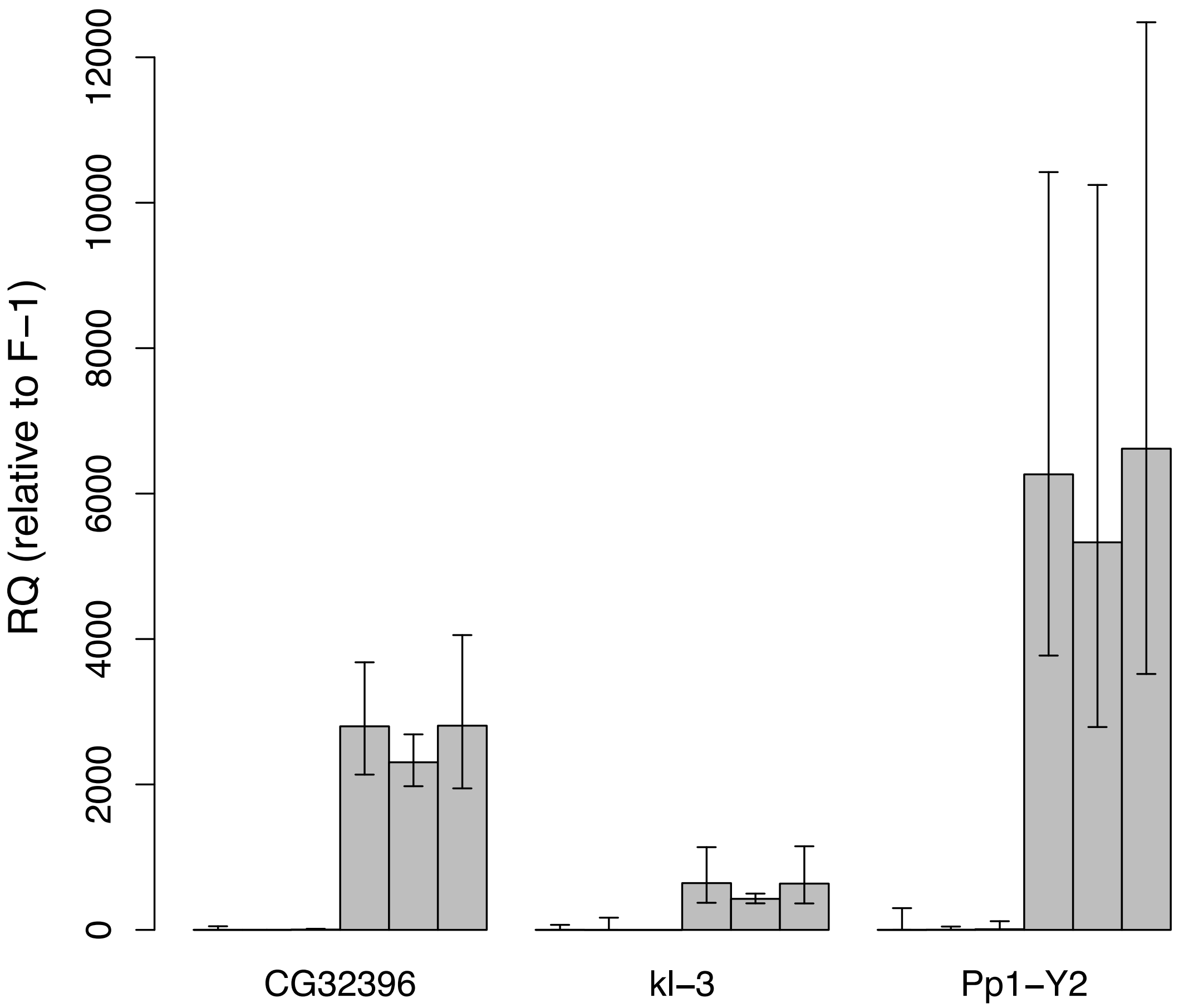

**Figure S7** Relative quantity of mRNA quantified by qRT-PCR in male and female larvae. In each category, the first three bars represent three independent female larvae sample (whole larva), each assayed with three technical replicates. The height of the bar represent the mean and the full range of RQ values were indicated by the error bars. The next three bars correspond to three independent male larvae assayed for the same gene. kl-3 and Pp1-Y2 are both located on the Y-chromosome and are known to have a testis-specific expression level. The RQ values were measured using RP49 gene as the internal control, and the first female larva sample (F-1) as the reference, whose RQ is set to one.

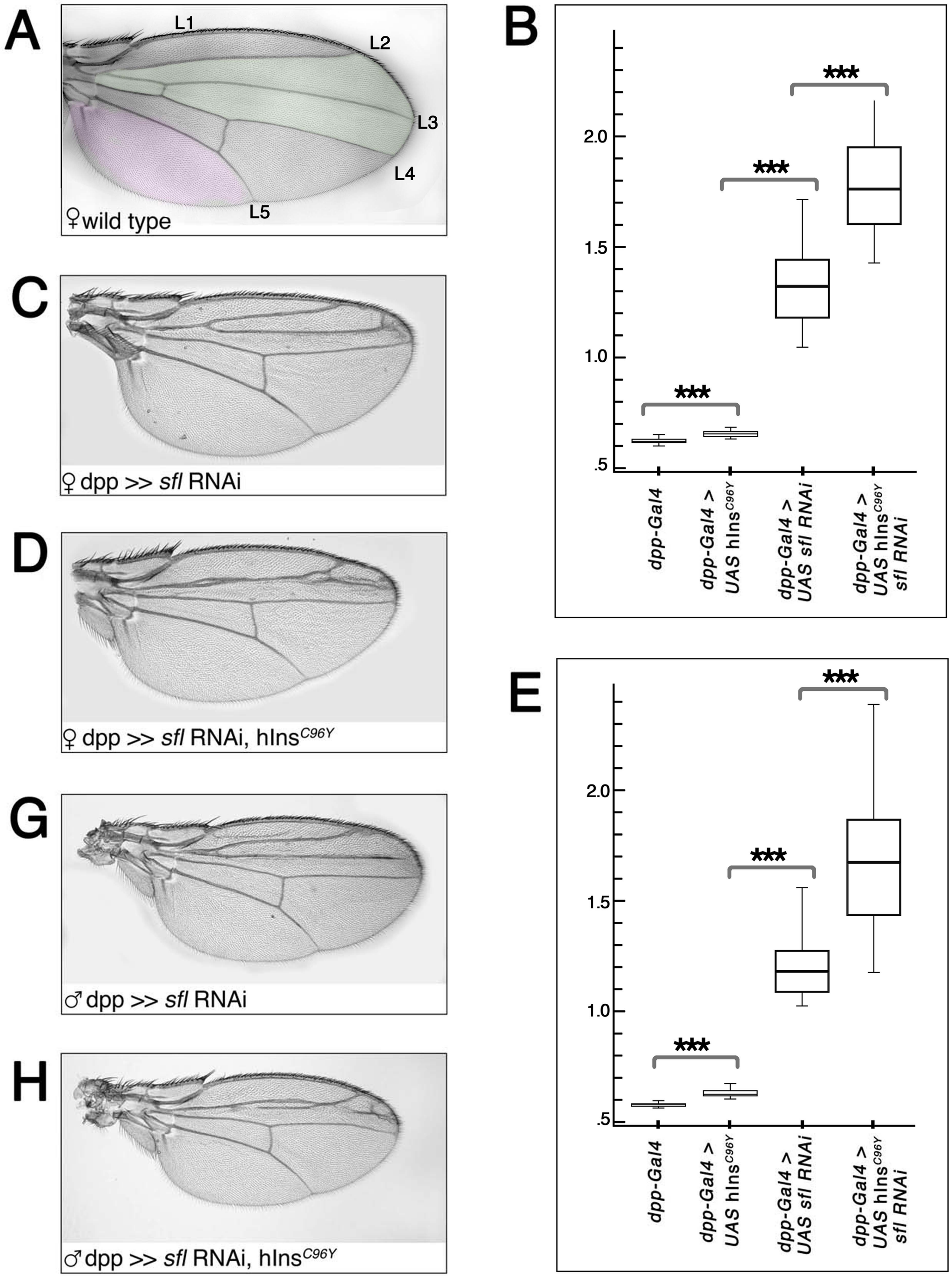

**Figure S8** Depletion of *sfl* by RNAi in the developing wing expressing hINS$^{C96Y}$ driven by dpp-Gal4. For both females and males, dpp >> hINS$^{C96Y}$ or Dpp >> *sfl* RNAi expression alone reduces wing area between the L2 and L4 longitudinal veins relative to the posterior-most sector of the wing (bordered by L5). This reduction is more severe in the *sfl* knockdown genotype than in the hINA$^{C96Y}$-expressing genotype. Co-expression of *sfl* RNAi and hINS $^{C96Y}$ by dpp-Gal4 results in the obliteration of the L3 vein and further relative reduction of the L2-L4 area.

(A): Wild type wing showing the measured regions of wing used to quantify the effects of both *sfl* RNAi and hINS$^{C96Y}$ expression in dpp-Gal4 domain (L3-L4 intervein sector). Quantification of the (B) female or (E) male wing phenotypes generated by transgenes dpp-Gal4; dpp-Gal4 >UAS-hINS$^{C96Y}$; (C, G) dpp-Gal4 >> UAS-*sfl* RNAi; and (D, H) dpp-Gal4 >>UAS-*sfl* RNAi; UAS-hINS$^{C96Y}$. The values represent the ratio of the third posterior cell (in pink color) divided by the L2-L4 intervein sector (in green color) wing area. ***, P < 0.001; Mann-Whitney U test.

Females: dpp-Gal4 (n= 15; Mean= 0.62), dpp-Gal4 >UAS-hINS$^{C96Y}$ (n= 15; Mean=0.65), dpp-Gal4 >> UAS-*sfl* RNAi (n= 23; Mean=1.3) and dpp-Gal4 >>UAS-*sfl* RNAi; UAS-hINS$^{C96Y}$ (n= 22; Mean=1.76).

Males: dpp-Gal4 (n= 15; Mean=0.59), dpp-Gal4 >UAS-hINS$^{C96Y}$ (n= 15; Mean=0.64), dpp-Gal4 >> UAS-*sfl* RNAi (n=23; Mean=1.2 ) and dpp-Gal4 >>UAS-*sfl* RNAi; UAS-hINS$^{C96Y}$ (n= 29; Mean=1.68).

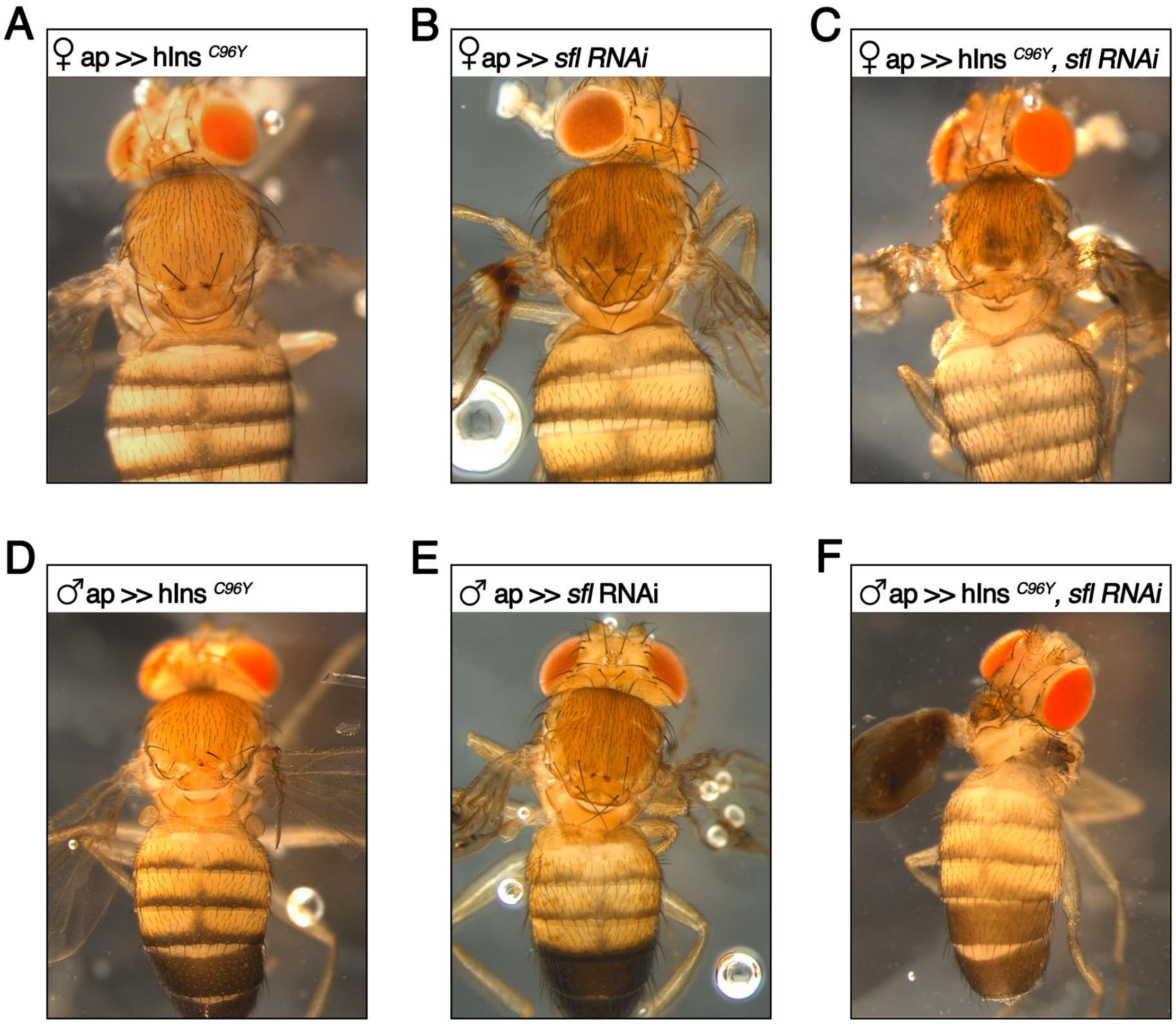

**Figure S9** Depletion of *sfl* by RNAi in the developing notum expressing $hINS^{C96Y}$ driven by ap-Gal4. For both females and males, ap > $hINS^{C96Y}$ or ap > *sfl* RNAi expression alone reduces notum area and causes loss of dorsal macrochaetae. Co-expression of *sfl* RNAi and $hINS^{C96Y}$ by ap-Gal4 results in greater destruction of the notum and macrochaetae in both sexes. However, in the male the notum and additional dorsal structures are obliterated and this phenotype is lethal.

ap-Gal4 > $hINS^{C96Y}$ (A) female and (D) male;

ap-Gal4 > *sfl* RNAi (B) female and (E) male;

ap-Gal4>> $hINS^{C96Y}$, *sfl* RNAi (C) female (F) male

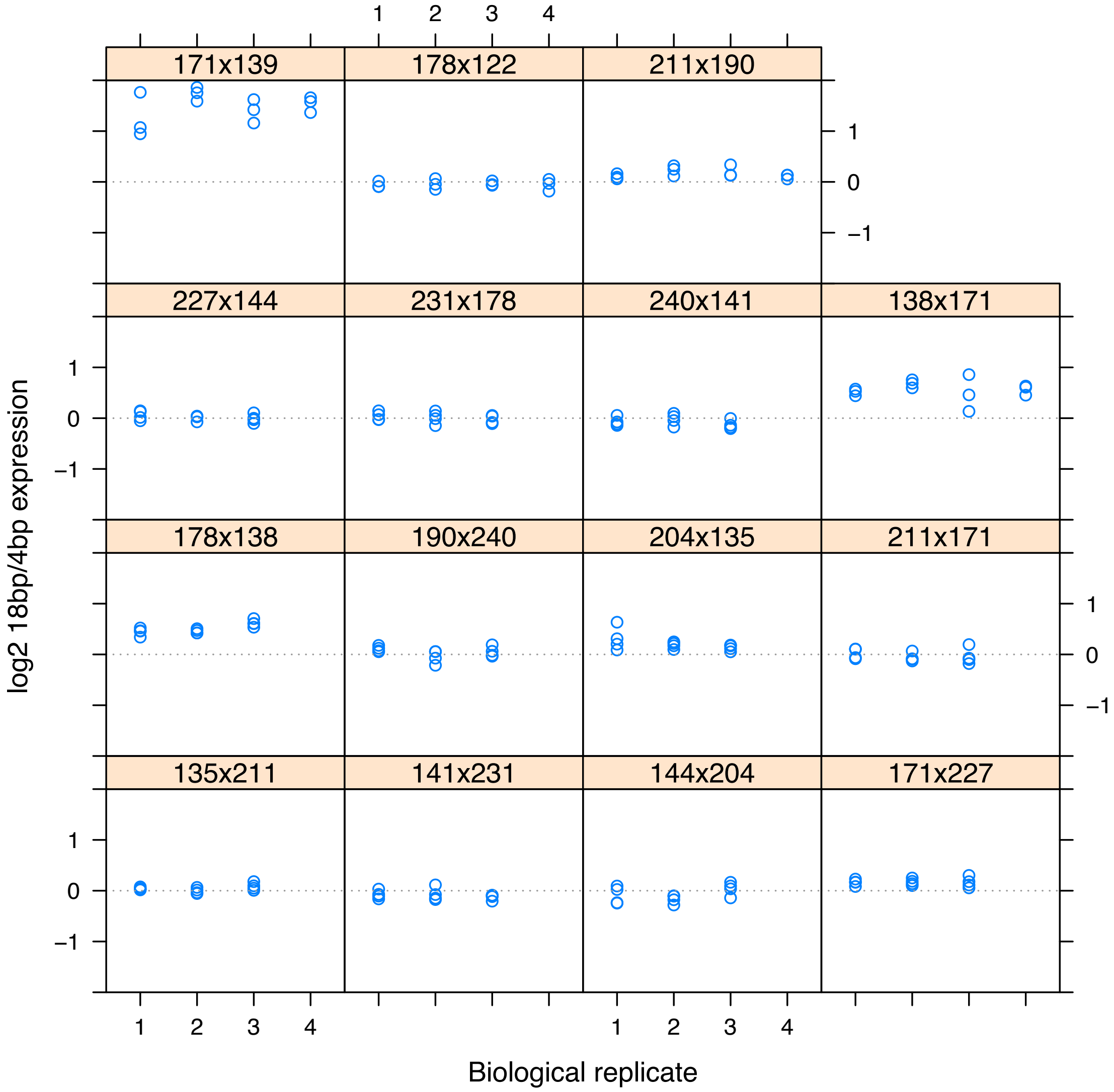

**Figure S10** log2 transformed ratios between transcript levels associated with 18bp/4bp alleles. The allele-specific expression ratios were measured in F1 hybrid individuals by pyro-sequencing, with three (or four) biological replicates and four (or three) pyro-technical replicates, to obtain a total of 12 measurements. In each of the 15 crosses, the technical replicates were plotted in a single column, with different columns representing the biological replicates. In the titles of each panel, the last three digits in the stock number were shown for lines used in the cross.

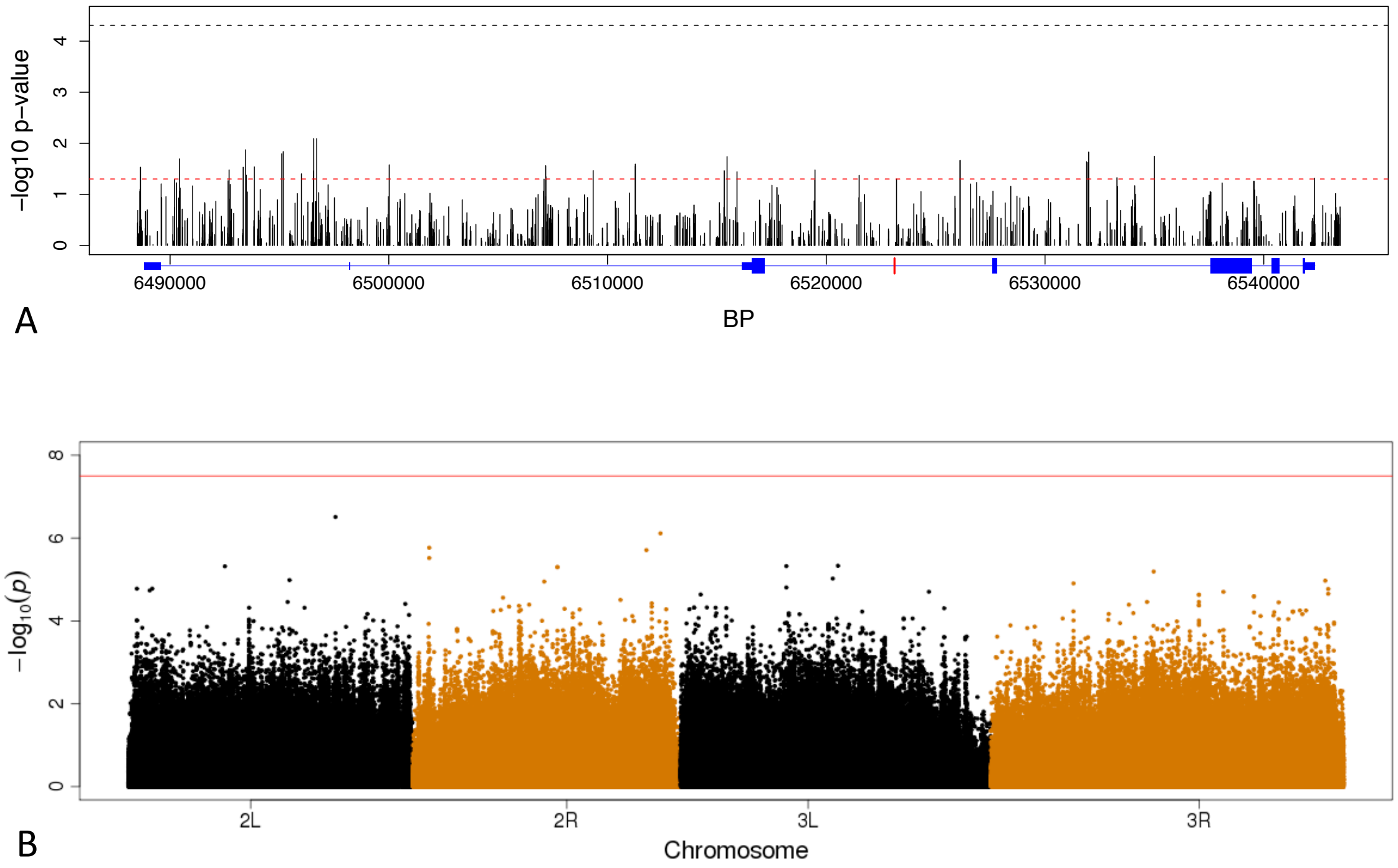

**Figure S11** Conditional regression analysis to detect additional SNPs associated with the phenotype of interest. (A) within the *sfl* locus; (B) all chromosomes. The intronic 18/4bp polymorphism in *sfl* is included in the linear model as a covariate. The two dotted lines in (A) correspond to a single test 0.05 level (red) and the multiple testing corrected 0.05 level using Bonferroni's method (blue). The red line in (B) represents the Bonferroni corrected 0.05 level.